\documentclass[journal,twoside,web]{ieeecolor}
\usepackage{generic}

\usepackage[english]{babel}
\usepackage[utf8]{inputenc}
\usepackage[T1]{fontenc}

\usepackage{microtype}
\usepackage{setspace}
\usepackage{blindtext}
\usepackage{siunitx}
\usepackage{comment}

\usepackage{titlecaps}
\Addlcwords{or the if a an of for and to -off off in not by via}
\usepackage[caption=false]{subfig}
\usepackage{graphicx}
\usepackage{color}
\usepackage{epsfig}
\graphicspath{{Images/}}
\usepackage{tikz}
\usepackage{pgfplots,pgfplotstable}
\usetikzlibrary{patterns,fit,matrix}
\usepackage{tikzscale}
\usepackage{scalerel}
\usetikzlibrary{arrows,
	patterns,
	plotmarks,
	svg.path,
	shapes.multipart}
\usepgfplotslibrary{fillbetween}
\pgfplotsset{compat=newest,
	compat/bar nodes=1.8,
	every axis/.append style={
		label style={font=\Large},
		tick label style={font=\large} 
	}
}
\tikzstyle{int}=[draw, fill=black!10, minimum size=5em,thick]
\tikzstyle{init} = [pin edge={to-,thick,black}]
\usepackage{booktabs}

\usepackage{array}
\let\labelindent\relax
\usepackage{enumitem}
\usepackage{setspace}

\usepackage{amsthm,amsmath,amssymb,amsfonts}
\usepackage{relsize}
\usepackage{nicefrac}
\usepackage{bbm}
\usepackage{mathtools}
\usepackage[mathscr]{eucal}
\usepackage[short,c2]{optidef}
\usepackage{dsfont}

\usepackage{url}
\usepackage[colorlinks]{hyperref}
\hypersetup{citecolor=blue,linkcolor=blue}
\usepackage[capitalize,nameinlink]{cleveref}

\usepackage{etoolbox}
\makeatletter
\@ifundefined{color@begingroup}%
{\let\color@begingroup\relax \let\color@endgroup\relax}{}%
\def\fix@ieeecolor@hbox#1{%
	\hbox{\color@begingroup#1\color@endgroup}}
\patchcmd\@makecaption{\hbox}{\fix@ieeecolor@hbox}{}{\FAILED}
\patchcmd\@makecaption{\hbox}{\fix@ieeecolor@hbox}{}{\FAILED}
\makeatother

\makeatletter
\newcommand{\orcid}[1]{\href{https://orcid.org/#1}{\includegraphics[scale=0.03]{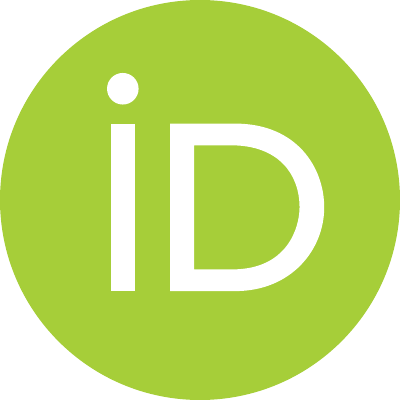}}} % link to ORCID

\newcommand{\journalVersion}[2]{#2\xspace} % use #1 for short, #2 for full version
\newcommand{\arxiv}[1]{#1}
\newcommand*\diff{\mathop{}\!\mathrm{d}}

\newcommand{\eg}{\emph{e.g.,}\xspace}
\newcommand{\ie}{\emph{i.e.,}\xspace}

\newcommand{\Real}[1]{ { {\mathbb R}^{#1} } }
\newcommand{\Realp}[1]{ { {\mathbb R}^{#1}_+ } }

\DeclarePairedDelimiter\floor{\lfloor}{\rfloor}

\newcommand{\pr}[1]{\mathbb{P}\ls#1\rs}

\newcommand{\mean}[1]{\mathbb{E}\left[#1\right]}

\newcommand{\one}[1][]{\mathds{1}_{#1}}

\newcommand{\e}{\mathrm{e}}

\renewcommand{\maxof}[2]{#1 \vee #2}
\renewcommand{\minof}[2]{#1 \wedge #2}

\newcommand{\norm}[2][]{\left\lVert#2\right\rVert_{#1}}

\theoremstyle{plain}
\newtheorem{thm}{Theorem}
\newtheorem{lemma}{Lemma}
\newtheorem{cor}{Corollary}
\newtheorem{prop}{Proposition}
\theoremstyle{definition}
\newtheorem{definition}{Definition}

\newtheorem{ass}{Assumption}
\theoremstyle{remark}
\newtheorem{rem}{Remark}

\newcommand{\lr}{\left(}
\newcommand{\rr}{\right)}
\newcommand{\ls}{\left[}
\newcommand{\rs}{\right]}
\newcommand{\lb}{\left\lbrace}
\newcommand{\rb}{\right\rbrace}

\newcommand{\agents}{\mathcal{V}}
\newcommand{\leg}{\mathcal{L}}
\newcommand{\mal}{\mathcal{M}}

\newcommand{\state}[2]{x_{#1}^{#2}}
\newcommand{\stateleg}[1]{x_{#1}^{\leg}}
\newcommand{\statemal}[1]{x_{#1}^{\mal}}
\newcommand{\neigh}[2][\@empty]{\mathcal{N}%
	\ifx\@empty#1 _{#2} \else _{#2}(#1) \fi}
\newcommand{\neighaug}[2][\@empty]{\neigh[#1]{#2}\cup\{#2\}}

\newcommand{\sumneighaug}[3][\@empty]{\sum_{#3\in\neighaug[#1]{#2}}}
\newcommand{\lam}[1]{\lambda_{#1}}
\newcommand{\statecontrleg}[1]{\bar{x}_{#1}^{\leg}}
\newcommand{\statecontrmal}[1]{\bar{x}_{#1}^{\mal}}
\newcommand{\statelegtil}[1]{y_{#1}^{\leg}}

\newcommand{\statemaltil}[1]{y_{#1}^{\mal}}

\newcommand{\stateerr}[2]{\tilde{x}_{#2}^{#1}}
\newcommand{\statelegerr}[2]{\tilde{x}_{#2}^{#1,\leg}}
\newcommand{\statelegtruess}{x_{\text{ss}}^{\leg,*}}
\newcommand{\statelegss}{x_{\text{ss}}^{\leg}}
\newcommand{\statecontrlegss}{\statecontrleg{\text{ss}}}
\newcommand{\statecontrmalss}{\statecontrmal{\text{ss}}}
\newcommand{\statemalerr}[2]{\tilde{x}_{#2}^{#1,\mal}}
\newcommand{\uleg}[2][\@empty]{u%
	\ifx\@empty#1 ^{\leg}\lr#2\rr \else _{#1}^{\leg} \fi}
\newcommand{\umal}[1][\@empty]{u%
	\ifx\@empty#1 ^{\mal} \else  _{\mal}\lr#1\rr\fi}

\newcommand{\Wleg}[1]{W_{#1}^{\leg}}
\newcommand{\Wmal}[1]{W_{#1}^{\mal}}
\newcommand{\Wlegtrue}{\overline{W}^{\leg}}
\newcommand{\Wlegerr}[2][\@empty]{\widetilde{W}%
	\ifx\@empty#1 _{#2}^{\leg} \else _{#2,#1}^{\leg} \fi}
\newcommand{\prodlam}[3]{\prod_{#1=#2}^{#3}(1-\lam{#1})}
\newcommand{\prodW}[3]{\prodlam{#1}{#2}{#3}\Wleg{#1}}
\newcommand{\prodWleg}[3]{\prod_{#1=#2}^{#3}\Wleg{#1}}
\newcommand{\prodWtrue}[3]{\prod_{#1=#2}^{#3}\Wlegtrue}
\newcommand{\prodWss}[3]{\prod_{#1=#2}^{#3}(1-\lam{#1})\Wlegtrue}
\newcommand{\prodlamW}[1]{\Pi_{#1}}
\newcommand{\prodinflam}[1]{\pi_{#1}}
\newcommand{\prodlamfin}[2]{\pi_{#1}^{#2}}
\newcommand{\tf}{T_\text{f}}
\newcommand{\tfmal}[1][\@empty]{T_\mal%
    \ifx\@empty#1 \else (#1) \fi}
\newcommand{\tfleg}[1][\@empty]{T_\leg%
	\ifx\@empty#1 \else (#1) \fi}
\newcommand{\dmax}{d_{\text{M}}}

\newcommand{\vmin}{v_\text{m}}

\newcommand{\Wautleg}[1]{W_{#1,\text{aut}}^{\leg}}
\newcommand{\Winleg}[1]{W_{#1,\text{in}}^{\leg}}
\newcommand{\Winmal}[2]{W_{#1,#2}^{\mal}}

\newcommand{\specrad}{\sigma}

\newcommand{\trust}[3]{\alpha_{#1#2}(#3)}
\newcommand{\trusthist}[3]{\beta_{#1#2}(#3)}
\newcommand{\meanleg}{E_\leg}
\newcommand{\meanmal}{E_\mal}
\newcommand{\eventcorrmal}[1]{\mathcal{E}_{\text{C,}\mal}(#1)}
\newcommand{\eventmisclmal}[1]{\mathcal{E}_{\text{M,}\mal}(#1)}
\newcommand{\eventcorrleg}[1]{\mathcal{E}_{\text{C,}\leg}(#1)}
\newcommand{\eventmisclleg}[1]{\mathcal{E}_{\text{M,}\leg}(#1)}

\addto\captionsenglish{\renewcommand{\figurename}{Fig.}}
\addto\captionsenglish{\renewcommand{\tablename}{TABLE}}

\newcommand{\blue}[1]{{\color{blue}#1}}

\newcommand{\revision}[1]{#1}

\newcommand{\linkToPdf}[1]{\href{#1}{\blue{(pdf)}}}
\newcommand{\linkToPpt}[1]{\href{#1}{\blue{(ppt)}}}
\newcommand{\linkToCode}[1]{\href{#1}{\blue{(code)}}}
\newcommand{\linkToWeb}[1]{\href{#1}{\blue{(web)}}}
\newcommand{\linkToVideo}[1]{\href{#1}{\blue{(video)}}}
\newcommand{\linkToMedia}[1]{\href{#1}{\blue{(media)}}}
\newcommand{\award}[1]{\xspace} % omit awards
\addto\extrasenglish{}
\addto\extrasenglish{}
\addto\extrasenglish{}

\Crefname{prop}{Proposition}{Propositions}
\Crefname{cor}{Corollary}{Corollaries}
\Crefname{ass}{Assumption}{Assumptions}
\Crefname{rem}{Remark}{Remarks}
\makeatother

\title{Confidence Boosts Trust-Based Resilience in \\ Cooperative Multi-Robot Systems}

\author{%
    Luca~Ballotta\textsuperscript{\orcid{0000-0002-6521-7142}},~\IEEEmembership{Member,~IEEE},
	\'Aron~V\'ek\'assy\textsuperscript{\orcid{0000-0002-6653-8300}},~\IEEEmembership{Graduate~Student~Member,~IEEE},
	Stephanie~Gil\textsuperscript{\orcid{0000-0002-4951-5350}},~\IEEEmembership{Member,~IEEE},
	and~Michal~Yemini\textsuperscript{\orcid{0000-0002-2087-1183}},~\IEEEmembership{Member,~IEEE}
    \thanks{This work has been partially supported
		by the Italian Ministry of Education, University and Research (MIUR) through the PRIN Project under grant 2017NS9FEY ``Realtime Control of 5G Wireless Networks,''
        by the ONR under grant N00014-21-1-2714,
        and by the AFOSR grant FA9550-22-1-0223.
        Views and opinions expressed in this work are of the authors and may not reflect those of the funding institutions.
    }%
	\thanks{Luca Ballotta is with the Department of Information Engineering, University of Padova, 35131 Padova, Italy
		(e-mail: luca.ballotta@unipd.it).}%
	\thanks{\'Aron V\'ek\'assy and Stephanie Gil are with the School of Engineering and Applied Sciences, Harvard University, Boston, MA 02138
		(e-mail: \{sgil, avekassy\}@g.harvard.edu).}%
	\thanks{Michal Yemini is with the Faculty of Engineering, Bar-Ilan University, Ramat-Gan 5290002 Israel
		(e-mail: michal.yemini@biu.ac.il).}% 
}

\begin{document}
	
    \bstctlcite{bib-options}
    \maketitle
    
    %!TEX ROOT = ../resilient_consensus_trust.tex

\begin{abstract}
	\boldmath
    \hypersetup{linkcolor=.}
	Wireless communication in multi-robot systems opens the door to cyberattacks that can disrupt safety and performance of collaborative robots.
	The physical channel supporting inter-robot communication offers an attractive opportunity to decouple the detection of malicious robots from data exchanged between legitimate robots.
	Yet,
	trustworthiness indications coming from physical channels are uncertain and must be handled with this in mind.
	In this paper,
	we propose a resilient consensus protocol wherein a parameter $\lam{t}$ accounts for how confident a robot is about the legitimacy of nearby robots that the physical channel indicates.
	Analytical results prove that our protocol achieves a resilient consensus as long as the communication graph without malicious robots supports consensus and trustworthiness indications are on average correct,
	without requiring redundant communication links such as $r$-robustness.
	Tuning $\lam{t}$ allows a designer to trade between near-optimal inter-robot coordination and quick task execution; see \autoref{fig:cover-slider}.
	This \emph{fundamental speed-deviation tradeoff} must be carefully balanced based on the task at hand.
	The effectiveness of our algorithm is numerically verified on autonomous platoons with spoofed vehicles.
\end{abstract}
    %!TEX ROOT = ../resilient_consensus_trust.tex

\section{Introduction}\label{sec:intro}

\begin{figure}
	\centering
	\includegraphics[width=\linewidth]{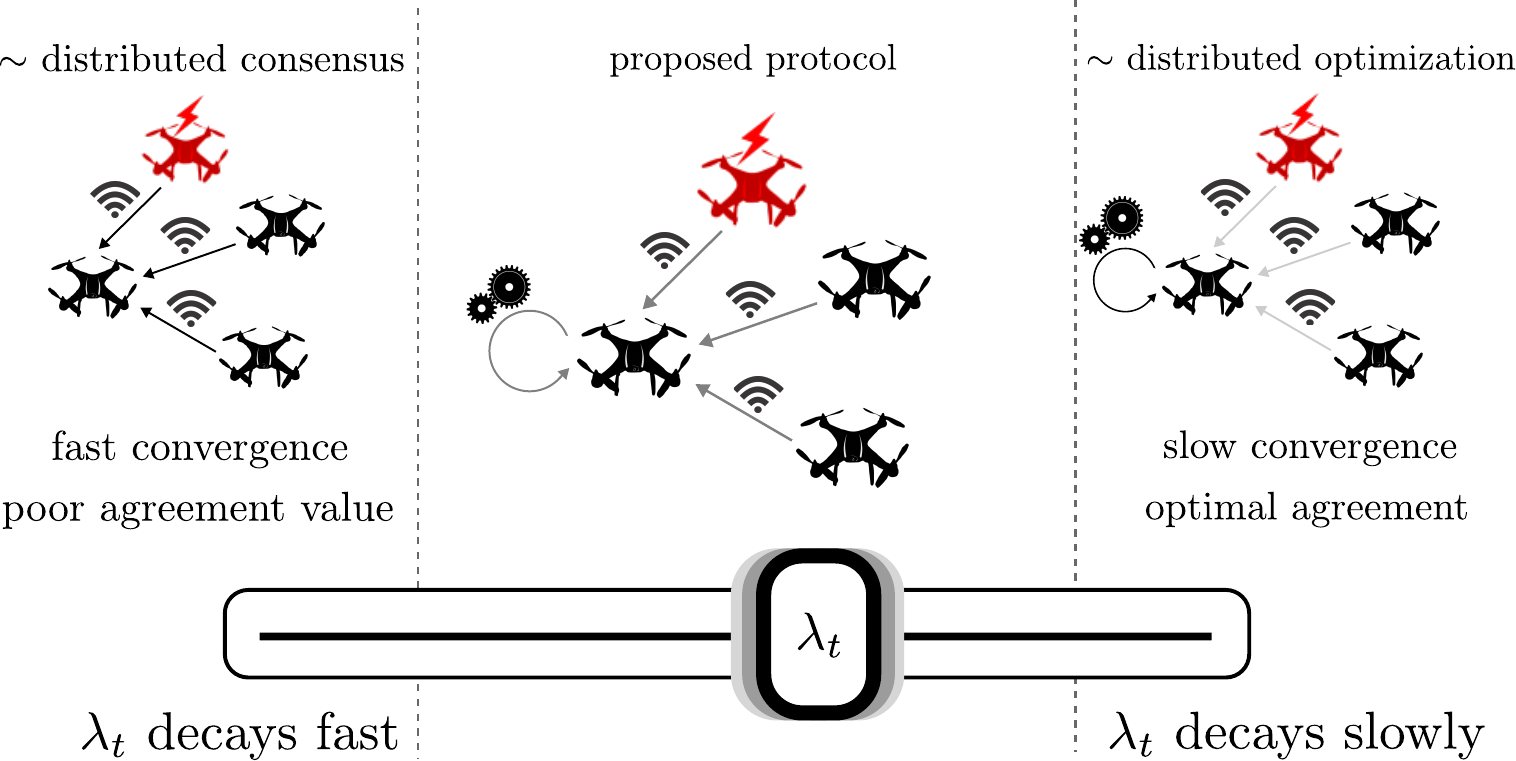}
	\caption{Our protocol lets robots {resiliently} cooperate {while they actively detect adversaries using stochastic trust observations derived from a physical channel,
			causing higher chance of misclassification during early iterations.}
		The {tunable} decay rate of the parameter $\lam{t}$ determines {how much each robot relies on messages received by its neighbors versus its own initial state,
			and allows a designer to} trade convergence speed for suboptimality of the consensus reached by robots.}
	\label{fig:cover-slider}
\end{figure}

\IEEEPARstart{M}{ulti}-robot systems are going to be key assets for transmission relay~\cite{Basturk22cn-multiRobotRelay}, underground and space exploration~\cite{Yliniemi14aimag-multiRobotExploration},
automated warehouses~\cite{Azadeh19ts-multiRobotWarehouse},
search-and-rescue~\cite{Baglioni24jirs-multiRobotSearchRescue},
and intelligent transportation~\cite{Eiras22tro-optimizationMotionPlanning}.
These tasks require smooth and reliable cooperation among robots to succeed.
At the same time,
they have to implement distributed control with local data exchange due to communication and computation constraints.

A fundamental gear in cooperative systems,
the consensus protocol lets robots agree on relevant quantities such as shared resources~\cite{Mahato23ras-taskAllocation},
perception models~\cite{Li22tro-resilientDistributedLearningRobots},
or relative locations~\cite{Zelazo15ijrr-rigidityMaintenance} through distributed averaging,
and underpins common distributed algorithms~\cite{Cortes17jcmsi-coordinatedControlMultiRobot,Shorinwa24ram-surveyDistributedOptimizationMultiRobot}.
However,
the consensus protocol is vulnerable to cyberattacks that can {pollute the messages transmitted from a robot to its neighbors over wireless network}.
Robots may agree on suboptimal values or miss consensus,
failing cooperative tasks and even threatening safety.
{Encryption imposes a computation and communication overhead which hampers real-time robot operations~\cite{fabris,fabris1}.
As such, lightweight resilient consensus strategies that directly robustify the distributed averaging are needed to enable effective multi-robot cooperation under adversarial influence.}
    	%!TEX ROOT = ../resilient_consensus_trust.tex

\subsection{Related work}
\label{sec:literature}

Resilience of multi-robot operation to {unknown malicious robots that interact with the rest of the team} has received a great deal of attention.
The survey~\cite{Prorok21arxiv-surveyResilienceMultiRobot} examines resilient strategies {based on average consensus} for multi-robot perception, planning, and control.
{In this context,
many works use filtering strategies that remove suspicious messages received based on their content. One of the most common filters is Weighted Mean Subsequence Reduced (W-MSR) which is guaranteed to ensure resilient consensus with a dense communication network}~\cite{LeBlanc13jsac-wmrs,Dibaji19acc-saba,Pirani23automatica-graphTheoreticResilientControl}.\footnote{
	{More precisely, it requires $r$-robust network graphs where $r$ relates to the number of malicious robots.
	This is a max cut-type graph-theoretic property which entails denser connectivity as $r$ increases.}
}
Early paper~\cite{Saldana18isdar-triangularNetworks} {fulfills this requirement by building triangular meshes on-the-fly}.
Follow-up works {focus on conditions to ensure consensus under increasing adversarial robots~\cite{Guerrero-Bonilla17ral-resilientFormations} and varying communication range~\cite{Guerrero-Bonilla19ral-resilientFormationsLattices}}.
{In a different spirit,}
paper~\cite{Cavorsi22rss-resilienceCBF} maintains dense connectivity via a control barrier function.
It requires the robots to estimate eigenvalues and eigenvectors of the communication Laplacian matrix,
hence it is sensitive to fast dynamics and non-resilient initial configurations.
{On the other hand,
most works assume a dense network and prove resilient behaviors with W-MSR.
For example,}
papers~\cite{Usevitch20tac-resilientLeaderFollower,Rezaee21automatica-resiliencyLeaderFollower,Santilli22tro-resilientContainment} study resilient leader-follower control,
Wang \emph{et al.}~\cite{Wang22tpds-resilientConsensusMobileMaliciousAgents} focus on mobile devices,
Dibaji \emph{et al.}~\cite{Dibaji19acc-saba} use buffers to discriminate adversaries based on all messages received,
and Wang \emph{et al.}~\cite{Wang20tcns-resilientConsensusEventBased} use event-based communication.
{However,
dense connectivity may not hold in large-scale robot teams,
its verification is NP-hard~\cite{LeBlanc13jsac-wmrs,Yi22cdc-approximateRRobustness},
and relies on a shared bound on the number of malicious robots whose imprecise estimate can degrade performance or fail consensus entirely}.
{In an effort to improve robustness,}
papers~\cite{Franceschelli17tac-medianConsensusRobustness,Shang22tcs-medianBasedResilientConsensus} investigate consensus to the median of robots' states instead of the average.
This method is robust to outliers but requires dense communication to mitigate malicious robots.

A few approaches do not presume dense interconnections.
References~\cite{Baras19med-trust,Bonagura23acc-resilientConsensusEvidence} filter neighbors with heuristic metrics of dissimilarity such as the Euclidean distance,
but assume that all malicious agents are eventually detected.
{Paper~\cite{Abbas18tcns-trustedNodes} pivots on a few ``trusted'' robots,}
which however may be expensive or infeasible to secure.
Work~\cite{Zhao18tsipn-resilientConsensusMobileDetectors} uses mobile robots that can listen to all transmissions and detect attackers by simply establishing contact, {which is practically challenging with scattered or resource-constrained robots.}
\revision{In~\cite{yuan21automatica-secureConsensus,Hadjicostis25tac-trustworthyConsensus}, agents exploit two-hop relaying of data packets to construct a binary trust assessment for each neighbor.
	However,
	reference~\cite{yuan21automatica-secureConsensus} assumes that the ID of every malicious agent is securely notified to all other agents upon detection,
	while~\cite{Hadjicostis25tac-trustworthyConsensus} assumes that the trust assessments converge to the true values in finite time.}

{All previous approaches neglect that robots feature physical components which can be used as a security watermark independent of received data}.
{We refer the reader to~\cite{Tsiamis20tac-stateSecrecyCodes,Gil17ar-spoofResilientMultiRobot,Xiong13icmcn-SecureArray,Pippin14icra-trustMultiRobotPatrolling,10436823,amin2024potentialreconfigurableintelligentsurface,10793103,10700780} for examples on how per-link trust information can be derived in practice,
for instance via the characteristics of the medium used for inter-robot communication~\cite{Tsiamis20tac-stateSecrecyCodes} or the angle-of-arrival of wireless signals~\cite{Xiong13icmcn-SecureArray,10436823}.}
On this positive note,
recent works leverage \emph{trust observations obtained from physical channels} to make consensus resilient without requiring dense connectivity or assuming that trust evaluations converge to the ground truth.
Reference~\cite{Yemini22tro-resilienceConsensusTrust} introduces a rule to weigh neighbors' messages based on random variables that describe trust and {formally establishes finite-time detection based on their statistics}.
Follow-up work~\cite{Yemini25tac-resilientDistributedOptim} extends~\cite{Yemini22tro-resilienceConsensusTrust} to resilient distributed optimization.
Papers~\cite{Mallmann-Trenn22tro-CrowdVetting,11162713} apply the physical trust framework to resilient multi-robot flocking with spoofed adversaries.
\revision{A comprehensive survey of attack models, computation of physics-based trust, and its use in distributed algorithms is provided in~\cite{Gil23-physicalityEnablesTrust}}.

This physical channel-based approach decouples the detection of adversaries from the data-driven cooperative task,
enabling formal performance guarantees under {standard assumptions on the adversary-free network graph, which are generally much milder than those required by data-driven filtering methods such as $r$-robustness}.
However,
the information collected from physical channels is uncertain,
partially hindering its usefulness if this is not properly accounted for.
While robots typically gain confidence in labeling neighbors as trustworthy or malicious as information is accrued,
individual transmissions are not reliable for such a classification.
This limitation generates a fundamental tradeoff between confidently classifying neighbors,
which takes nonnegligible time,
and the fast decision-making robotic tasks require.
        %!TEX ROOT = ../resilient_consensus_trust.tex

\subsection{Contribution}
\label{sec:contribution}

In this paper,
we design a novel protocol for resilient multi-robot collaboration with unknown adversaries.
We draw inspiration from recent works on resilient consensus that respectively use trust information from physical channels~\cite{Yemini22tro-resilienceConsensusTrust} and the {Friedkin-Johnsen (FJ)} model~\cite{Ballotta24tac-competitionCollaboration},
and propose a best-of-both-worlds approach integrating \textit{trust observations} robots obtain from the channel and {the \textit{confidence} robots have about trustworthiness of their neighbors.
Our protocol anchors the robots to their initial state through a decaying weight $\lam{t}$ that reflects how confident they feel about classification of neighbors,
as the stochastic nature of trust information sources requires accruing observations over a time horizon before confidently (dis)trusting a robot.
This countermeasure avoids that legitimate robots rely too much on malicious messages they misclassify as trustworthy.}
The confidence parameter $\lam{t}$ generates a fundamental tradeoff between deviation and speed which is depicted in \autoref{fig:cover-slider}.
If $\lam{t}$ decays slowly,
the robots precisely converge to the nominal adversary-free consensus but after long time,
possibly hindering rapid decision-making;
if $\lam{t}$\linebreak decays fast,
the robots agree to a suboptimal value in short time.
{This dichotomy is reminiscent of a classic tradeoff between convergence speed and accuracy in distributed optimization.
	In fact, the parameter $\lam{t}$ can be interpreted as the stepsize of noisy distributed gradient descent~\cite{Akgun25lcss-FJisDGD}.}

Our hybrid approach overcomes two practical limitations of previous works~\cite{Yemini22tro-resilienceConsensusTrust,Ballotta24tac-competitionCollaboration}.
Robots do not use an idle observation window before starting the consensus algorithm as done in~\cite{Yemini22tro-resilienceConsensusTrust},
{enabling fast collaborative and resilient decision-making}.
Paper~\cite{Ballotta24tac-competitionCollaboration} uses a constant confidence parameter $\lam{t}\equiv\lambda$ that prevents a final agreement,
whereas we use a vanishing $\lam{t}$ which makes legitimate robots agree eventually.
{Our protocol is formally backed by~\cite{Ballotta24lcss-FJDiminishingCompetition} which proves that it recovers the nominal consensus without adversaries,
ensuring no loss of performance under ideal conditions.}

This paper significantly extends the conference version~\cite{Ballotta24acc-confidenceTrust} by improving the bounds on deviation from nominal consensus,
analyzing the convergence rate,
and reporting extensive simulations.
{Compared to~\cite{Ballotta24lcss-FJDiminishingCompetition},
including malicious robots and trust observations makes the consensus dynamics stochastic,
significantly changing its analysis and convergence results.}

\subsubsection*{Organization}
\autoref{sec:system-model} introduces the multi-robot system setup, including nominal cooperative objective and attack model.
We present our resilient consensus protocol in \autoref{sec:resilient-protocol}.
We analytically characterize it in \autoref{sec:performance-analysis},
including convergence to a consensus among legitimate robots (\autoref{sec:convergence}),
deviation from the nominal consensus value (\autoref{sec:deviation}),
and quantification of the convergence speed (\autoref{sec:convergence-rate}).
Numerical experiments show the effectiveness of our approach in \autoref{sec:simulations}.
We discuss current limitations and directions of future work in \autoref{sec:limitations} and conclude in \autoref{sec:conclusions}.
    %!TEX ROOT = ../resilient_consensus_trust.tex

\section{System Setup and Objective}\label{sec:system-model}

\begin{table}
	\caption{Main symbols used in this paper. If a star ($^*$) precedes a description, the quantity associated with it is unknown in practice and used only for the sake of analysis.}
	\label{tab:symbols}
	\centering
	\small
	\begin{tabular}{ll}
		\toprule
		$\agents,\leg,\mal$	& Sets of all, $^*$legitimate, $^*$malicious robots\\
		$N,L,M$	& Number of all, $^*$legitimate, $^*$malicious robots\\
		$\state{t}{i}$	& State of robot $i$ at time $t$\\
		$\state{t}{}$, $\stateleg{t}$, $\statemal{t}$	& States of all, legitimate, malicious robots\\
		$\statelegtruess$	& $^*$Nominal consensus value of legitimate robots\\
		$\statelegss$	& Final consensus value of legitimate robots\\
		$\stateerr{i}{t}$	& $^*$Distance of state $\state{t}{i}$ from $\statelegtruess$\\
		$\eta$	& Maximal magnitude of a robot's state\\
		$\trust{i}{j}{t}$	& Trust observation of robot $i$ about $j$ at time $t$\\
		$\trusthist{i}{j}{t}$	& Aggregated trust of robot $i$ about $j$ till time $t$\\
		$\neigh{i}$	& Set of in-neighbors of robot $i$\\
		$\neigh[t]{i}$	& Set of trusted in-neighbors of robot $i$ at time $t$\\
		$\Wlegtrue$	& $^*$Nominal communication weight matrix \\
		$w_{ij}(t)$	& Weight robot $i$ assigns to $j$ at time $t$\\
		$\Wleg{t}$, $\Wmal{t}$	& Weights assigned to legitimate, malicious robots\\
		$\lam{t}$	& Confidence parameter at time $t$\\
		$\meanleg,\meanmal$	& $^*$Expected values of trust variables minus $\nicefrac{1}{2}$\\
		$\tf,\tfleg,\tfmal$	& $^*$Final misclassification times of $\agents$, $\leg$, $\mal$\\
		\bottomrule
	\end{tabular}
\end{table}

In this section we present the multi-robot system model,
{the attack model,}
and the collaborative consensus task at hand.
{The main notation is summarized in \autoref{tab:symbols}.}

\subsubsection*{Multi-robot system}
Consider $N$ robots equipped with scalar-valued states.
Our algorithm can handle multi-dimensional states,
but this choice avoids burdensome notation.
{We tag each robot with a unique number $i \in \agents \doteq \{1,\dots,N\}$.}
The state of robot $i$ at time $t\in \mathbb{N}\cup\{0\}$ is $\state{t}{i}\in\Real{}$ and the vector $\state{t}{}\in\Real{N}$ stacks all states $\{\state{t}{i}\}_{i\in\agents}$.
Robots communicate through a communication network modeled as a time-invariant directed graph $\mathcal{G} = (\agents, \mathcal{E})$.
Each element $e = (i,j) \in \mathcal{E}$ indicates that robot $j$ can transmit data to robot $i$ through a direct link.
We denote the in-neighbors of robot $i$ in the network $\mathcal{G}$ by $\neigh{i} \doteq \{j\in\agents\setminus\{i\} : (i,j)\in\mathcal{E}\}$ and the maximal in-degree of all robots by $\dmax\doteq\max_{i\in\agents}|\neigh{i}|$.

\subsubsection*{{Attack model}}
{In the network,
	$L$ \emph{legitimate robots} truthfully follow a designated protocol aimed at achieving a collaborative goal,
	while $M=N-L$ are \textit{malicious robots}.
	We respectively partition them in sets $\leg\subseteq\agents$ and $\mal\subseteq\agents$ such that $\mal\cap\leg=\emptyset$ and denote their stacked states by $\stateleg{t}\in\Real{L}$ and $\statemal{t}\in\Real{M}$.
	Malicious robots are fixed, \ie $\mal$ never changes.
	\revision{Attacks occur through peer-to-peer communication, as follows.
		Let $x_{t}^{i,\textnormal{nom}}$ denote the nominal state that malicious robot $i\in\mal$ would have if it followed the designated protocol.
		During operations, it transmits (potentially) different values $\state{t}{i}\neq x_{t}^{i,\textnormal{nom}}$ to its neighbors.\footnote{
			\revision{Our algorithm does not require that malicious robots broadcast the same value to all neighbors, but we use this convention for the sake of readability.}}}
	Legitimate robots do not know if or which neighbors are malicious.
	\emph{The objective of this work is to develop a distributed protocol that recovers, as close as possible, the outcome of a nominal cooperative protocol despite interactions with unknown malicious robots}.
	In addition to exchanging states $\state{t}{j}$, transmissions from robot $j$ to $i$ are tagged by a signal $\alpha_{ij}(t)$. Its significance and mathematical model will be clarified later.
	\revision{No robot $j\in\agents$ can manipulate the signal $\trust{i}{j}{t}$ which tags its own transmissions.
	This is motivated by physical layer security and, more broadly, physics-based frameworks whereby a \emph{receiving} robot $i$ computes $\trust{i}{j}{t}$ by comparing real-time physical features with a model~\cite{Gil17ar-spoofResilientMultiRobot,Mallmann-Trenn22tro-CrowdVetting,Gil23-physicalityEnablesTrust}.}
\subsubsection*{Consensus task}
The legitimate robots aim to agree on a common state,
{\ie to achieve a consensus}.
The nominal consensus value is determined by the initial states $\stateleg{0}$ and the {nominal communication network that connects only the legitimate robots}.
Define the nominal communication weight matrix $\Wlegtrue\in\Real{L\times L}$ with uniform weights
\begin{equation}\label{eq:weights-true}
	\ls\Wlegtrue\rs_{i j}=
	\begin{cases}
		\frac{1}{\left|\neigh{i}\cap\leg\right|+1} & \text{if } j \in (\neigh{i}\cap\leg)\cup\{i\} \\ 
		0 & \text{otherwise}.
	\end{cases}
\end{equation}
Ideally,
the legitimate robots should disregard messages sent by malicious robots and run the nominal consensus protocol
\begin{equation}\label{eq:consensus-protocol-nominal}\tag{NOM}
	\stateleg{t+1} = \Wlegtrue\stateleg{t}, \quad t\ge0.
\end{equation}
The following standard assumption is necessary to reach consensus {even in the ideal scenario without malicious robots}.
\begin{ass}[Primitive matrix~\cite{Olfati-Saber07pieee-consensus}]\label{lem:W-primitive}
	Matrix $\Wlegtrue$ is primitive; there exists a stochastic vector $v$ such that $\lim_{t\to\infty}(\Wlegtrue)^t=\one v^\top$. 
\end{ass}
The vector $v$ is called the Perron-vector of $\Wlegtrue$ and its $i$th element quantifies how much robot $i$ influences the final consensus.
If \cref{lem:W-primitive} holds,
protocol~\eqref{eq:consensus-protocol-nominal} drives all robots' states to  $\statelegtruess \doteq \lim_{t\to\infty}\state{t}{i}= v^\top\stateleg{0}$,
for all $i\in\leg$.

{However,
as the malicious robots are unknown,}
the legitimate robots cannot implement the weights~\eqref{eq:weights-true} and run the protocol~\eqref{eq:consensus-protocol-nominal}.
In the next section,
we propose a resilient consensus protocol to ideally recover the final outcome of~\eqref{eq:consensus-protocol-nominal} notwithstanding malicious robots.
\revision{\cref{lem:W-primitive} is our only requirement on the network;
	we do not assume a dense graph or bound the number of malicious robots interfering with legitimate ones.}
Instead,
we use trust observations that each legitimate robot independently obtains from the physical channel to \revision{\emph{decentralize the detection of malicious neighbors,} removing the need for redundant links.}
    %!TEX ROOT = ../resilient_consensus_trust.tex

\section{Resilient Consensus Protocol}\label{sec:resilient-protocol}

In this work,
we propose the following distributed protocol to be implemented by each legitimate robot $i\in\leg$
\begin{equation}\label{eq:update-rule-regular}\tag{RES}
	\state{t+1}{i} = \lam{t}\state{0}{i} + \lr1-\lam{t}\rr\sumneighaug{i}{j}w_{ij}(t)\state{t}{j}, {\quad t\ge0}.
\end{equation}
Communication weights $ w_{ij}(t) \in[0,1] $ are computed based on \textit{trust observations} that robot $i$ has accrued about neighbor $j$ till time $t$.
Trusted neighbors are given positive weights and the others zero weight,
in the attempt to reconstruct the nominal weights~\eqref{eq:weights-true}.
We formally define trust and explain how robots compute communication weights in \autoref{sec:trust}.
Parameter $\lam{t} \in\ [0,1] $ accounts for how uncertain robot $i$ feels about its decision to trust, or not, its neighbors after $t$ updates.
Equivalently,  
$1-\lam{t}$ captures the \textit{confidence} of robot $i$ about the trustworthiness of its neighbors,
and serves to mitigate potential mistakes in the assignment of communication weights.
For the sake of brevity,
we call $\lam{t}$ the confidence parameter in the following,
and describe it in \autoref{sec:confidence}.

The parameter $\lam{t}$ is new w.r.t. previous works on trust-based resilience and in this paper we characterize the impact of this ``confidence'' term on mitigating malicious robots when~\eqref{eq:update-rule-regular} starts from time $t=0$.
{This approach contrasts with the observation window $[0,T_0]$ used in paper~\cite{Yemini22tro-resilienceConsensusTrust} which delays the consensus protocol by $T_0>0$ time steps}.
In fact,
with our formalism,
the algorithm in~\cite{Yemini22tro-resilienceConsensusTrust} is the special case of~\eqref{eq:update-rule-regular} where $\lam{t}$ is a step function that is equal to $1$ till time $T_0$ and equal to zero afterwards.
Hence,
our approach where $\lam{t}$ can be arbitrarily tuned is more general.

{\begin{rem}[Ideal network]\label{rem:ideal-communication}
	Robot $i$ instantaneously receives the state $\state{t}{j}$ from robot $j\in\neigh{i}$ with no error.
	However,
	motivated by trust variables from physical channels,
	weights $w_{ij}(t)$ are stochastic,
	as explained next.
\end{rem}}
        %!TEX ROOT = ../resilient_consensus_trust.tex

\subsection{Embedding Trust: The Communication Weights}\label{sec:trust}

We assume that each transmission from robot $j$ to $i$ is tagged with an observation $\trust{i}{j}{t}\in[0,1]$ of a random variable $\alpha_{ij}$.

\begin{definition}[Trust variable $\alpha_{ij}$]\label{def:alpha}
	For every $i\in\leg$ and $j\in\neigh{i}$,
	the random variable $\alpha_{ij}$ {is supported on $[0,1]$} and represents the probability that robot $j$ is a trustworthy neighbor of robot $i$.
	{Observations $\trust{i}{j}{t}$ of $\alpha_{ij}$ are available at all times $t\ge0$ and are independent across all robot pairs and time instants.}
\end{definition}

{Statistically independent trust observations are supported in the literature, such as the discrete memoryless channel model.}
A realization $\trust{i}{j}{t}$ of $\alpha_{ij}$ contains useful information if the legitimacy of the transmission can be thresholded.
We assume that $\trust{i}{j}{t}>\nicefrac{1}{2}$ indicates a legitimate transmission and $\trust{i}{j}{t}<\nicefrac{1}{2}$ a malicious transmission in a stochastic sense (miscommunications are possible).
If $\trust{i}{j}{t}=\nicefrac{1}{2}$,
the transmission at time $t$ contains no useful information.
{We do not assume a specific distribution of trust variables.}

We draw inspiration from~\cite{Yemini22tro-resilienceConsensusTrust} and choose the weights $w_{ij}(t)$ in~\eqref{eq:update-rule-regular} according to the observation history.
The aggregate trust robot $ i\in\leg $ has accrued about robot $ j\in\neigh{i} $ at time $t$ is
\begin{equation}\label{eq:trust-history}
	\trusthist{i}{j}{t}\doteq\sum_{s=0}^t\lr\trust{i}{j}{s} - \dfrac{1}{2}\rr {= \trusthist{i}{j}{t-1} + \trust{i}{j}{t} - \dfrac{1}{2}.}
\end{equation}
Robot $i$ computes its \textit{trusted neighborhood} at time $ t $
\begin{equation}\label{eq:trusted-neighborhood}
	\neigh[t]{i}\doteq\lb j \in \neigh{i}:\trusthist{i}{j}{t} \geq 0 \rb
\end{equation}
and assigns communication weights at each time $t$
\begin{equation}\label{eq:weights-trust}
	w_{i j}(t)= 
	\begin{cases}
		\frac{1}{\left|\neigh[t]{i}\right|+1} & \text{if } j \in \neigh[t]{i} \cup \{i\}, \\ 
		0 & \text{otherwise}.
	\end{cases}
\end{equation}
The trusted neighborhood $\neigh[t]{i}$ is designed to reconstruct the set $\neigh{i}\cap\leg$ leveraging trust information collected by robot $i$,
and rule~\eqref{eq:weights-trust} aims to recover nominal weights~\eqref{eq:weights-true}.
        %!TEX ROOT = ../resilient_consensus_trust.tex

\subsection{Weighing Trust: The Confidence Parameter}\label{sec:confidence}

Intuitively,
robot $i$ accrues knowledge about the trustworthiness of its neighbors as more trust-tagged transmissions have been received.
This intuition can be formalized by upper bounding the probability of misclassifying a neighbor,
which we do with \cref{lem:final-misclassification} below.
First, we require a basic assumption.
\begin{ass}[Trust observations are informative]\label{ass:trust-meaningful}
	For any robot $i\in\leg$,
	define expectations $\meanleg\doteq\mean{\alpha_{ij}} - \nicefrac{1}{2}$ for $j\in\leg\cap\neigh{i}$ and $\meanmal\doteq\mean{\alpha_{ij}} - \nicefrac{1}{2}$ for $j\in\mal\cap\neigh{i}$.
	Then,
	it holds $\meanleg>0$ and $\meanmal<0$.
\end{ass}
{\cref{ass:trust-meaningful} implies that the detection implemented via trust observations is meaningful, namely,
	the $\trust{i}{j}{t}$ correctly distinguish legitimate from malicious robots, \emph{on average}.
	If this was not the case,
	trust variables would be useless as their classification accuracy would be worse than flipping a coin.
	This fact is formally stated in~\cite{Yemini22tro-resilienceConsensusTrust} where it is shown that a robot can be misclassified with positive probability forever if \cref{ass:trust-meaningful} fails, making convergence analysis trivial.
	The closer $\meanleg$ (resp., $\meanmal$) is to $1/2$ (resp., $-1/2$),
	the more accurate the information provided on average by observations $\trust{i}{j}{t}$.
	The values of $\meanleg$ and $\meanmal$ are not needed for design and we use them only to analyze the performance of~\eqref{eq:update-rule-regular}.}
\begin{lemma}[Decaying misclassification probability~\cite{Yemini22tro-resilienceConsensusTrust}]\label{lem:misclassification-probability}
	Let \cref{ass:trust-meaningful} hold.
	Then,
	it holds
	\begin{equation}\label{eq:prob-misclassification}
		\begin{aligned}
			\pr{\trusthist{i}{j}{t} < 0} &\le \e^{-2\meanleg^2(t+1)} &&\forall i\in\leg, j\in\neigh{i}\cap\leg\\
			\pr{\trusthist{i}{j}{t} \ge 0} &\le \e^{-2\meanmal^2(t+1)} &&\forall i\in\leg, j\in\neigh{i}\cap\mal.
		\end{aligned}
	\end{equation}
\end{lemma}
Towards the next results,
we recall that an event occurs \emph{almost surely} if it has probability $1$.\arxiv{\footnote{
	Formally,w
	a probability measure is defined over a $\sigma$-algebra $\mathcal{F}\in2^\Omega$ of the sample space $\Omega$.
	In this work,
	the sample space collects all possible realizations (observations) $\alpha_{ij}(t)$ of the trust variables.
}
Equivalently,
the event not happening has zero chance.}

\begin{cor}[Final misclassification time]
	\label{lem:final-misclassification}
	There exist finite times $\tfmal\ge0$ and $\tf\ge\tfmal$ such that {all malicious robots are correctly classified} for $t\ge\tfmal$ and {all robots are correctly classified} for $t\ge\tf$,
	almost surely.
    \begin{proof}
        The result of $\tf$ is proven in \cite{Yemini22tro-resilienceConsensusTrust}. 
        Because $\tf$ (resp., $\tfmal$) refers to correct classification of all (resp. malicious) robots,
        it holds $\tf\ge\tfmal$ since legitimate robots can be misclassified after all malicious robots are correctly classified.
    \end{proof}
\end{cor}

\cref{lem:final-misclassification} establishes successful classification of all robots after some unknown, but finite, time $\tf$.
We will rely on this fact for asymptotic analysis.
Conversely, a higher chance of misclassification exists during early iterations by \cref{lem:misclassification-probability}.
Hence,
we introduce the \emph{decreasing} parameter $ \lam{t} $ to enhance resilience.
{Since $1-\lam{t} \gtrsim 0$ for small $ t $,
robot $i$ assigns small weight to the values received by its trusted neighbors $\neigh[t]{i}$,
making early updates nearly insensitive to attacks from misclassified malicious neighbors in $\neigh[t]{i}\cap\mal$.
On the other hand,
\cref{lem:misclassification-probability} establishes fast decaying probability of misclassification and so we set $\lam{t}\gtrsim0$ for larger $t$,
when trusted neighbors are likely legitimate,
to recover a consensus.}

\begin{rem}[Trust statistics]\label{rem:trust-non-iid}
	Our characterization of trust variables $\alpha_{ij}$ can be relaxed. \cref{lem:final-misclassification} holds with link-dependent expectations satisfying \cref{ass:trust-meaningful} at infinity or with time correlations under a submartingale condition on per-link trust observations; see~\cite{Yemini25tac-resilientDistributedOptim}.
	However, we stick to i.i.d. trust variables for the sake of exposition.
\end{rem}

\begin{rem}[Related work]
    Paper~\cite{Khatana23tsipn-noiseResilientConsensus} uses a similar approach to improve PushSum with noise but neglects attacks.
\end{rem}

\arxiv{%
    \subsubsection*{Resilience by trust and confidence}
    The update rule~\eqref{eq:update-rule-regular} is designed to jointly leverage the two key concepts of \emph{trust} and \emph{confidence},
    which are independently used in previous works.
    
    The papers~\cite{Yemini22tro-resilienceConsensusTrust,Yemini22cdc-resilientDistributedOptim,Yemini25tac-resilientDistributedOptim}
    use physics-based trust observations to help legitimate robots decide which neighbors they should rely on.
    At each time-step,
    the robot either trusts a neighbor or not,
    but it does not scale the weights given to trusted neighbors relatively by how confident it is on the decision. 
    Furthermore, 
    in~\cite{Yemini22tro-resilienceConsensusTrust} the deviation from nominal consensus is strongly tied to an initial observation window $[0,T_0]$ where the robots do not update their states and only collect trust observations to choose wisely which neighbors to trust in the first update round.
    Choosing the value of $T_0$ is nontrivial when the total number of rounds is not known in advance,
    and the method requires accurate synchronization.
    In contrast,
    we introduce the parameter $\lam{t}$ to capture the confidence that a robot has about the trustworthiness of its neighbors,
    proposing a softer approach to the clear-cut observation window in~\cite{Yemini22tro-resilienceConsensusTrust}.
    
    The use of $\lam{t}$ draws inspiration from previous work~\cite{Ballotta22cdc-competitionCollaboration,Ballotta24tac-competitionCollaboration}
    that uses the Friedkin-Johnsen model~\cite{Friedkin90jms-fjModel} to achieve resilient average consensus,
    intended as the minimization of the mean square deviation.
    However,
    these references adopt a \textit{constant} parameter $\lambda_t\equiv\lambda$ that prevents consensus to happen.
    In contrast,
    in this work we use the physical channel to obtain trust information independently of the data and make the competition-based rule able to recover a consensus,
    which is relevant to several control and robotic applications.%
}
    %!TEX ROOT = ../resilient_consensus_trust.tex

\section{Analysis}\label{sec:performance-analysis}

In this section,
we analytically assess performance of~\eqref{eq:update-rule-regular}.
In \autoref{sec:convergence},
we prove that the legitimate robots reach a consensus.
Then,
we quantify performance along two axes.
In \autoref{sec:deviation},
we upper bound the steady-state deviation from the nominal consensus without malicious robots.
This quantifies suboptimality of~\eqref{eq:update-rule-regular} if it runs long enough.
Then,
we quantify the finite-time deviation in \autoref{sec:convergence-rate} which indicates how fast the protocol converges.
As the analysis reveals,
a tension exists between deviation and speed,
influenced by how fast the parameter $\lam{t}$ decays.
Before diving into the analysis,
we introduce a few convenient notations.

Let $W_t\in\Real{L\times N}$ denote the matrix with weights~\eqref{eq:weights-trust},
\ie $[W_t]_{ij} = w_{ij}(t)$,
and partition the weight matrix into weights given to legitimate robots,
$\Wleg{t}$,
and to malicious robots,
$\Wmal{t}$
\begin{equation}
	W_t = \ls\begin{array}{c|c}
		\Wleg{t} &\Wmal{t}
	\end{array}\rs, \quad \Wleg{t}\in\Real{L\times L}, \Wmal{t}\in\Real{L\times M}.
\end{equation}
This partition is done for the sake of analysis only, since the legitimate robots do not know the adversaries' identities.
The protocol~\eqref{eq:update-rule-regular} can be compactly written in vector form as
\begin{equation}\label{eq:dynamics}
	\stateleg{t+1}	= \lam{t}\stateleg{0} + (1-\lam{t})\begin{bmatrix}
			\Wleg{t} & \Wmal{t}
		\end{bmatrix}\begin{bmatrix}
			\stateleg{t}\\
			\statemal{t}
		\end{bmatrix}.
\end{equation}
The state of legitimate robots $\stateleg{t}$ embeds messages transmitted by both legitimate and malicious robots.
To study performance,
it is convenient to set these two contributions apart, as we will do next.
Define the following transition matrices at time $t$
\begin{subequations}\label{eq:leg-state-contr}	
	\begin{align}
		\Wautleg{t} &\doteq \prodW{k}{0}{t-1}\label{eq:Waut}\\
		\Winleg{t} &\doteq \sum_{k=0}^{t-1} \lr\prodW{s}{k+1}{t-1}\rr\lam{k}\label{eq:Winleg}\\
		\Winmal{k}{t} &\doteq \lr\prodW{s}{k+1}{t-1}\rr (1-\lam{k})\Wmal{k}\label{eq:Winmal}
	\end{align}
	that respectively represent the consensus-type weighted averaging of legitimate robots' states,
	the effect of the constant legitimate input $\stateleg{0}$,
	and that of malicious inputs $\statemal{k}$,
    all at time $t$.
	We define the state contributions due to (messages transmitted by) legitimate and malicious robots at time $t$ respectively as 
	\begin{align}
		\statecontrleg{t} &\doteq \lr\Wautleg{t} + \Winleg{t}\rr\stateleg{0} \label{eq:leg-state-contr-leg}\\
		\statecontrmal{t} &\doteq \sum_{k=0}^{t-1}\Winmal{k}{t}\statemal{k}.\label{eq:leg-state-contr-mal}
	\end{align}
	\arxiv{The contribution $\statecontrleg{t}$ (resp., $\statecontrmal{t}$) incorporates only state values transmitted by legitimate (resp., malicious) robots.}
\end{subequations}
Subbing $\statecontrleg{t}$ and $\statecontrmal{t}$ defined in~\eqref{eq:leg-state-contr-leg}--\eqref{eq:leg-state-contr-mal} into~\eqref{eq:dynamics} yields
\begin{equation}\label{eq:state-contributions}
	\stateleg{t} = \statecontrleg{t} + \statecontrmal{t}.
\end{equation}
Motivated by practical considerations such as physical saturation of robot actuators,
we assume that the states communicated by malicious robots are bounded.
If adversaries send implausible or infeasible large values,
they can be immediately detected by thresholding.
We use the same bound on initial conditions for the sake of notation,
but this does not affect the analysis.
\begin{ass}[State bound]\label{ass:initial-state-bound}
	There exists $\eta \in\Realp{}$ such that $\max_{i\in\leg}|\state{0}{i}|\le\eta$
	and $\max_{i\in\mal,t\ge0}|\state{t}{i}|\le\eta$.
\end{ass}
        %!TEX ROOT = ../resilient_consensus_trust.tex

\subsection{Convergence to Consensus}\label{sec:convergence}

We now make the primary step that proves our proposed approach meaningful.
\begin{prop}[Protocol~\eqref{eq:update-rule-regular} achieves a consensus]\label{prop:consensus}
	Let \cref{lem:W-primitive,ass:trust-meaningful} hold and $\lim_{t\rightarrow\infty}\lam{t}=0$.
	Then,
	there exists scalar $\statelegss\in\Real{}$ such that,
    almost surely,
	\begin{equation}\label{eq:consensus}
		 \lim_{t\rightarrow\infty}\stateleg{t} = \statelegss\one.
	\end{equation}
	\begin{proof}
		See \cref{app:consensus}.
	\end{proof}
\end{prop}

{\cref{prop:consensus} reveals that the legitimate robots achieve a consensus under~\eqref{eq:update-rule-regular} if $\lam{t}$ vanishes.
On the other hand,
if $\lam{t}$ is ultimately bounded away from zero,
consensus does not generally happen~\cite{Ballotta24lcss-FJDiminishingCompetition}.
Therefore, we assume that vanishing $\lambda_t$ can be imposed by a system designer.}

\begin{ass}[Vanishing $\lam{t}$]\label{ass:vanishing-lam}
	It holds that $\lim_{t\rightarrow\infty}\lam{t}=0$.
\end{ass}

While all choices of {vanishing} sequences $\{\lambda_t\}_{t\geq0}$ lead to convergence almost surely, we show in the next sections that the \textit{specific} choice of sequence  $\lam{t}$ affects the \textit{performance} of the protocol~\eqref{eq:update-rule-regular} with respect to deviation and speed.
        %!TEX ROOT = ../resilient_consensus_trust.tex

\subsection{Steady-State Deviation From Nominal Consensus}\label{sec:deviation}

Given that legitimate robots achieve a consensus,
we assess how far the trajectory of~\eqref{eq:update-rule-regular} deviates from that of the nominal protocol~\eqref{eq:consensus-protocol-nominal}.
The deviation of robot $i$ at time $t$ is
\begin{equation}\label{eq:deviation}
	\stateerr{i}{t} \doteq \left|\state{t}{i} - \statelegtruess\right| = \left|\ls\stateleg{t} - \one\statelegtruess\rs_i\right|.
\end{equation}
To quantify the worst-case deviation from the nominal consensus,
one may seek bounds $\epsilon>0$ and $\delta>0$ such that
\begin{equation}\label{eq:max-ss-deviation-limsup}
	\pr{\max_{i\in\leg}\limsup_{t\rightarrow\infty}\stateerr{i}{t} > \epsilon} < \delta,
\end{equation}
namely the chance that any legitimate robot's final state $\state{\infty}{i}$ is more than $\epsilon$ distant from the nominal consensus $\statelegtruess$ is at most $\delta$.
However,
\cref{prop:consensus} states that legitimate robots almost surely reach a consensus.
This allows us to formally neglect the maximization over $\leg$,
that is trivial almost surely,
and to compute the chance of~\eqref{eq:max-ss-deviation-limsup} for every $i$ as follows
\begin{multline}\label{eq:limsup-prob-partition}
	\pr{\limsup_{t\rightarrow\infty}\:\stateerr{i}{t} > \epsilon}
	=\pr{\limsup_{t\rightarrow\infty}\:\stateerr{i}{t} > \epsilon \cap \tf <\infty}\\
	+ \pr{\limsup_{t\rightarrow\infty}\:\stateerr{i}{t} > \epsilon \cap \tf =\infty}.
\end{multline}
According to \cref{lem:final-misclassification},
it holds $\pr{\tf <\infty}=1$.
Moreover,
the proof of \cref{prop:consensus} shows that $\lim_{t\rightarrow\infty}\stateerr{i}{t}$ exists for all $i\in\leg$ if $\tf$ is finite.
Therefore,
we simplify~\eqref{eq:limsup-prob-partition} as
\begin{equation}\label{eq:lim}
	\begin{aligned}
		\pr{\limsup_{t\rightarrow\infty}\:\stateerr{i}{t+1} > \epsilon}
		&=\pr{\limsup_{t\rightarrow\infty}\:\stateerr{i}{t} > \epsilon \cap \tf <\infty}\\
        &=\pr{\limsup_{t\rightarrow\infty}\:\stateerr{i}{t} > \epsilon \;|\; \tf <\infty}\\
		&=\pr{\lim_{t\rightarrow\infty}\stateerr{i}{t} > \epsilon \;|\; \tf <\infty}.
	\end{aligned}
\end{equation}
By virtue of~\eqref{eq:lim},
in the following we assess the final deviation from nominal consensus by computing $\delta(\epsilon)$ such that
\begin{equation}\label{eq:max-ss-deviation}
	\pr{\lim_{t\rightarrow\infty}\stateerr{i}{t} > \epsilon \;|\; \tf <\infty} < \delta(\epsilon).
\end{equation}
For the sake of readability only, 
hereafter we omit the conditioning event and use compact notations such as $\pr{\lim_{t\rightarrow\infty}\stateerr{i}{t} > \epsilon}$ in place of~\eqref{eq:max-ss-deviation} whenever we assume $\tf<\infty$ such that the limit exists.

\arxiv{Evaluating~\eqref{eq:max-ss-deviation} helps achieve insight to design the parameter $\lam{t}$.}
To analytically compute $\delta(\epsilon)$,
it is convenient to separately assess the state contributions of legitimate and malicious robots and then combine their respective bounds.
Formally,
we write
\begin{equation}\label{eq:deviation-split}
	\begin{aligned}
		\stateerr{i}{t}	&= \left|\ls\stateleg{t} - \one\statelegtruess\rs_i\right|\\
								&\stackrel{\eqref{eq:leg-state-contr}}{=} \left|\ls\statecontrleg{t} + \statecontrmal{t} - \one\statelegtruess\rs_i\right| \le\statelegerr{i}{t} + \statemalerr{i}{t},
	\end{aligned}
\end{equation}
where,
defining the matrix
\begin{equation}\label{eq:state-leg-err-matrix}
	\Wlegerr{t} \doteq \Wautleg{t} + \Winleg{t}	- \one v^\top,
\end{equation}
the \emph{deviation due to legitimate robots} is given by
\begin{equation}\label{eq:state-leg-error}
	\statelegerr{i}{t} = \left|\ls \statecontrleg{t} - \one\statelegtruess \rs_i\right| = \left|\ls \Wlegerr{t}\stateleg{0} \rs_i\right|
\end{equation}
while the \emph{deviation due to malicious robots} is simply the magnitude of their contribution to the legitimate robots' states
\begin{equation}\label{eq:state-mal-error}
	\statemalerr{i}{t} \doteq \left|\ls \statecontrmal{t}\rs_i\right|.
\end{equation}
This splitting facilitates the analysis and reflects the impact of malicious robots on the system behavior.
Since the consensus task involves only legitimate robots as explained in \autoref{sec:system-model},
term~\eqref{eq:state-leg-error} should ideally vanish.
Conversely,
term~\eqref{eq:state-mal-error} means that messages sent by malicious robots should be discarded.

To make analysis more tractable and \revision{explicitly show how the decay rate of $\lam{t}$ affects the final deviation from nominal value,}
in this subsection we choose $\lam{t}$ as
\begin{equation}\label{eq:lambda}
	\lam{t} = c\e^{-\gamma t}, \qquad c\in(0,1), \ \gamma > 0.
\end{equation}
This choice satisfies \cref{ass:vanishing-lam}.
Intuitively,
if $\gamma$ is small,
the legitimate robots do not fully collaborate with trusted neighbors for many iterations,
which helps resilience when the trust variables are highly uncertain.
Conversely,
large values of $\gamma$ practically turn~\eqref{eq:update-rule-regular} into a consensus protocol after a few iterations and are suitable when the true weights $\Wlegtrue$ are quickly recovered.
\revision{The choice~\eqref{eq:lambda} is also motivated by the exponentially decreasing misclassification probabilities in \cref{lem:misclassification-probability} since, ideally,
	the decay rate $\gamma$ should be tuned to the decay rates of the exponents in~\eqref{eq:prob-misclassification}.}
Further,
\revision{simulations suggest that the analysis applies to other designs of $\lam{t}$; see \autoref{sec:simulations}}.
            %!TEX ROOT = ../resilient_consensus_trust.tex

\subsubsection{Deviation due to legitimate robots}

We first upper bound the expectation of the deviation term in~\eqref{eq:state-leg-error} at steady state.

\begin{prop}[Deviation due to legitimate robots]\label{lem:deviation-leg}
	Define
        \iffalse
        \begin{equation}
            s(\gamma)\doteq-\frac{1}{\gamma}-\frac{\ln(1-ce^{-\gamma})}{\gamma}\cdot \frac{1-ce^{-\gamma}}{ce^{-\gamma}}\label{eq:s}
        \end{equation}
        \fi
	\begin{equation}\label{eq:z}
        z(\gamma;k) \doteq -\frac{1}{\gamma}-\frac{\ln(1-ce^{-\gamma(k+1)})}{\gamma}\cdot \frac{1-ce^{-\gamma(k+1)}}{ce^{-\gamma(k+1)}}.
	\end{equation}
	Under \cref{lem:W-primitive,ass:trust-meaningful,ass:vanishing-lam,ass:initial-state-bound} and $\tf<\infty$,
	the deviation from nominal consensus due to legitimate robots is upper bounded as
	\begin{equation}\label{eq:leg-err-prob-bound}
		\mean{\lim_{t\rightarrow\infty}\statelegerr{i}{t}} \le \eta\uleg[]{} \quad \forall i\in\leg
	\end{equation}
	where $\uleg[]{} \doteq \min\{\uleg[\text{aut}]{} + \uleg[\text{in}]{},2\}$ and
	\begin{align}
		\uleg[\text{aut}]{} &\doteq 2\e^{z(\gamma;0)}\lr1-\lr\dfrac{1}{\dmax+1}\rr^{\mean{\tf}}\rr\label{eq:u1}\\
		\uleg[\text{in}]{} &\doteq 2 \mean{\sum_{k=0}^{\tf-2} \e^{z(\gamma;k+1)}\lam{k}\lr1-\dfrac{1}{\lr\dmax+1\rr^{\tf-k-1}}\rr}\!.\label{eq:u2}
	\end{align}
	\begin{proof}
		We derive the bound in two parts respectively associated with two components of $\statelegerr{i}{t}$.
		From~\cite{Ballotta24lcss-FJDiminishingCompetition}, running~\eqref{eq:update-rule-regular} with true weights (\ie $\Wlegtrue$ for legitimate and zeros for malicious robots) leads to the nominal consensus,
		or equivalently
		\begin{equation}
			\prodWss{k}{0}{\infty} + \sum_{k=0}^{\infty} \lr\prodlam{s}{k+1}{\infty}\Wlegtrue\rr\lam{k} = \one v^\top.
		\end{equation}
		In light of this,
		we split the matrix~\eqref{eq:state-leg-err-matrix} that accounts for the deviation due to legitimate robots as
		\begin{equation}\label{eq:Wt}
			\Wlegerr{t} = \Wlegerr[\text{aut}]{t} + \Wlegerr[\text{in}]{t},
		\end{equation}
		where we highlight that the matrix associated with the deviation from the nominal (autonomous) consensus dynamics is
		\begin{equation}\label{eq:leg-err-contribution-lambda}
			\Wlegerr[\text{aut}]{t} \doteq \Wautleg{t} - \prod_{k=0}^{\infty}(1-\lambda_k)\Wlegtrue
		\end{equation}
		and the one associated with the legitimate input $\{\lam{t}\stateleg{0}\}_{t\ge0}$ is
		\begin{equation}\label{eq:leg-err-contribution-input}
			\Wlegerr[\text{in}]{t} \doteq \Winleg{t} - \sum_{k=0}^{\infty} \lr\prodlam{s}{k+1}{\infty}\Wlegtrue\rr\lam{k}.
		\end{equation}
		The same arguments in the proof of \cref{prop:consensus} show that both the two deviation terms respectively associated with $\Wlegerr[\text{aut}]{t}$ and $\Wlegerr[\text{in}]{t}$ converge to a consensus if $\tf<\infty$.
		Applying the triangle inequality to~\eqref{eq:state-leg-error} with~\eqref{eq:Wt} and assuming that the limits exist yields
		\begin{equation}\label{eq:Wt-tr-ineq}
			\lim_{t\rightarrow\infty}\statelegerr{i}{t} \le \lim_{t\rightarrow\infty}\left| \ls \Wlegerr[\text{aut}]{t}\stateleg{0}\rs_i \right| + \lim_{t\rightarrow\infty}\left| \ls \Wlegerr[\text{in}]{t}\stateleg{0}\rs_i \right|.
		\end{equation}
		By linearity of expectation,
		we get
		\begin{multline}\label{eq:leg-err-markov-limit}
			\mean{\lim_{t\rightarrow\infty}\statelegerr{i}{t}} \le \mean{\lim_{t\rightarrow\infty}\left| \ls \Wlegerr[\text{aut}]{t}\stateleg{0}\rs_i \right|} \\ + \mean{\lim_{t\rightarrow\infty}\left| \ls \Wlegerr[\text{in}]{t}\stateleg{0}\rs_i \right|}.
		\end{multline}
		We separately upper bound the two expectations above as
		\begin{gather}
			\label{eq:deviation-leg-one-bound}
			\mean{\lim_{t\rightarrow\infty}\left| \ls \Wlegerr[\text{aut}]{t}\stateleg{0}\rs_i \right|} \le \eta\uleg[\text{aut}]{}\\
			\label{eq:deviation-leg-two-bound}
			\mean{\lim_{t\rightarrow\infty}\left| \ls \Wlegerr[\text{in}]{t}\stateleg{0}\rs_i \right|} \le \eta \uleg[\text{in}]{}
		\end{gather}
		with $\uleg[\text{aut}]{}$ and $\uleg[\text{in}]{}$ defined in~\eqref{eq:u1} and~\eqref{eq:u2}.
		The detailed derivation of bounds~\eqref{eq:deviation-leg-one-bound} and~\eqref{eq:deviation-leg-two-bound} is provided in \cref{app:proof-deviation-leg}.
	\end{proof}
\end{prop}

\begin{figure}
	\centering
	\includegraphics[width=.6\linewidth]{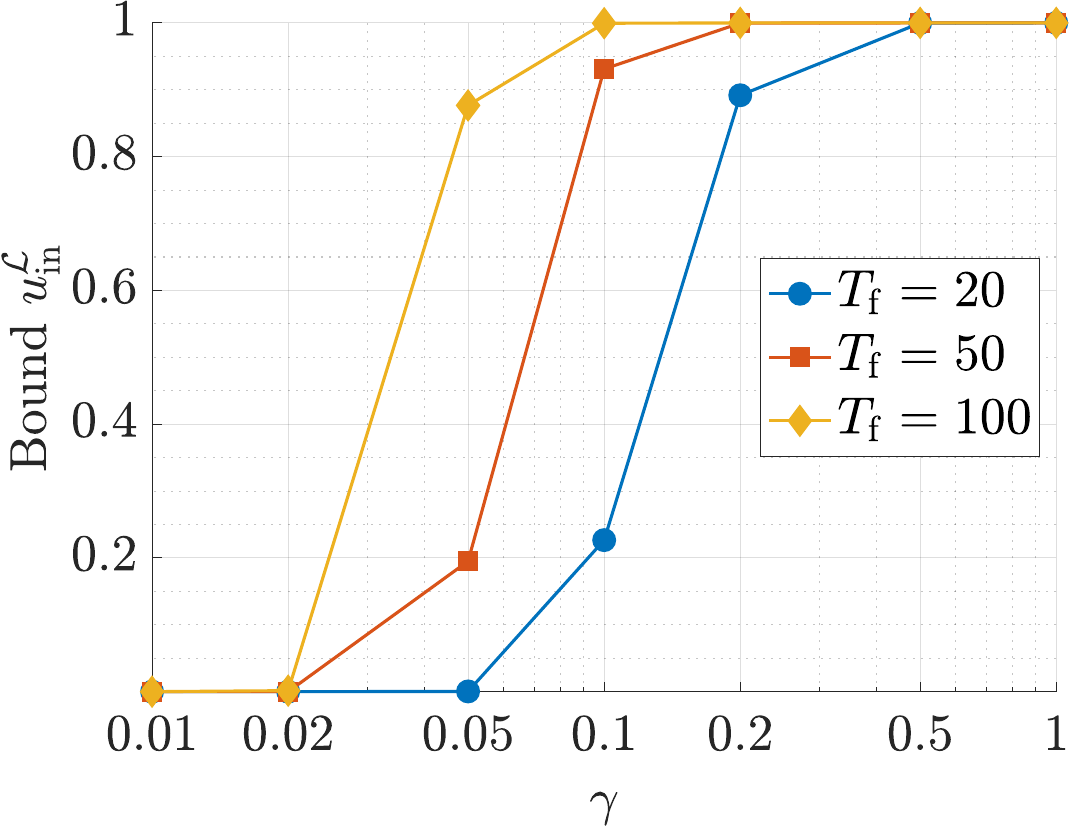}
	\caption{Upper bound $\uleg[]{}$ in~\cref{lem:deviation-leg} on expected deviation due to legitimate robots for several values of $\tf$ with $\dmax = 9$ and $\lam{t}=0.9\e^{-\gamma t}$.}
	\label{fig:u_leg}
\end{figure}

The bound $\uleg[]{}$ increases with $\tf$.
This is intuitive because a large $\tf$ means that some legitimate neighbors are not trusted for long time,
losing useful information.
Conversely,
although the term $\uleg[\text{aut}]{}$ in~\eqref{eq:u1} increases with $\gamma$,
the cumbersome expression of the overall bound $\uleg[]{}$ prevents us from analyzing its monotonicity w.r.t. $\gamma$.
Numerical evaluation suggests that $\uleg[]{}$ increases with $\gamma$ analogously to $\uleg[\text{aut}]{}$,
as visible in \autoref{fig:u_leg},
such that a slow decay of $\lam{t}$ reduces the deviation.
This is also intuitive because,
if $\lam{t}$ decays slowly,
the legitimate robots do not rely much on incoming messages for a long time,
mitigating all misclassification.
This reminds of the strategy in~\cite{Yemini22tro-resilienceConsensusTrust} where consensus starts at time $T_0$ and a larger $T_0$ reduces the deviation.
The key difference is that $\lam{t}<1 \, \forall t>0$ in~\eqref{eq:update-rule-regular},
allowing the legitimate robots to update their states in $\stateleg{t}$ from the beginning without waiting for $T_0$ time-steps.
            %!TEX ROOT = ../resilient_consensus_trust.tex

\subsubsection{Deviation due to malicious robots}\label{sec:deviation-malicious}

The following result quantifies the harmful effects of malicious robots.
\begin{prop}[Deviation due to malicious robots]\label{lem:deviation-mal}
	Define the following quantities
	\begin{gather}
		D_\mal \doteq \sum_{i\in\leg}|\mal\cap\neigh{i}|\label{eq:D}\\
		\begin{multlined}\label{eq:zeta}
			\zeta \doteq \frac{1}{\e^{2\meanmal^2}-1}\lr\frac{1}{1-\e^{-2\meanmal^2}}-\frac{1}{1-\e^{-4\meanmal^2}}\rr\\
			-\dfrac{c\left(1+\e^{-\gamma}\right)}{\e^{2\meanmal^2}-\e^{-\gamma}}\lr\frac{1}{1-\e^{-2\meanmal^2}}-\frac{1}{1-\e^{-4\meanmal^2-\gamma}}\rr\\
			+\frac{c^2\e^{-\gamma}}{\e^{2\meanmal^2}-{\e^{-2\gamma}}}\left(\frac{1}{1-\e^{-2\meanmal^2}}-\frac{1}{1-\e^{-4\meanmal^2-2\gamma}}\right).
		\end{multlined}
	\end{gather}
	Under \cref{ass:trust-meaningful,ass:vanishing-lam,ass:initial-state-bound} and $\tf<\infty$,
	the deviation from nominal consensus due to malicious robots is upper bounded as
	\begin{equation}\label{eq:state-mal-error-bound-exp}
		\mean{\lim_{t\rightarrow\infty}\statemalerr{i}{t}} \le \eta\umal \quad \forall i\in\leg
	\end{equation}
	where
	\begin{equation}
		\umal = \dfrac{D_\mal^2}{2}\zeta.
	\end{equation}
	\begin{proof}
		See \cref{app:proof-deviation-mal}.
	\end{proof}
\end{prop}
The bound $\umal{}$ in~\eqref{eq:state-mal-error-bound-exp} increases with $\gamma$.\footnote{
	The term $\zeta$ is a sum of terms all increasing with $\gamma$; see \cref{app:proof-deviation-mal}.
}
This means that,
if $\lam{t}$ decays slowly (\ie the regime where $\gamma$ is small),
the legitimate robots reduce the weight of messages received from malicious neighbors in times where detection may be unreliable due to a small sample size of trust observations,
which reduces the final deviation.
Since $\umal{}$ decreases with $\meanmal^2$ and $\meanmal<0$ by \cref{ass:trust-meaningful},
more uncertain trust observations associated with malicious transmissions prolong their misclassifications and increase the expected deviation.

\arxiv{\begin{rem}[Tightening bound~\eqref{eq:state-mal-error-bound-exp}]
		\label{rem:tighter-bound-mal-agents}
		The bound on deviation due to malicious agents can be improved by using tighter bounds on the weights $[\Wmal{t}]_{ij}$ and on the probability of correct classification time $\pr{\tfmal=k}$.
		The resulting bound can be computed in closed form but amounts to a huge,
		cumbersome expression.
		The derivation of this bound,
		which may be used for numerical evaluation,
		is provided in~\cref{app:tighter-bound-mal-agents}.
\end{rem}}
            %!TEX ROOT = ../resilient_consensus_trust.tex

\subsubsection{Bound on deviation}\label{sec:deviation-total}

The overall bound on the deviation from nominal consensus can be computed by merging the two bounds obtained for legitimate and malicious robots' contributions.
The following result quantifies how far from nominal consensus the legitimate robots eventually get.
\begin{thm}[Deviation from nominal consensus with~\eqref{eq:update-rule-regular}]\label{thm:bound}
	Under \cref{lem:W-primitive,ass:trust-meaningful,ass:vanishing-lam,ass:initial-state-bound},
	the deviation from nominal consensus is upper bounded as
	\begin{equation}\label{eq:deviation-bound-prob}
		\pr{\lim_{t\rightarrow\infty}\stateerr{i}{t} > \epsilon} \le \dfrac{\eta}{\epsilon}\lr\uleg[]{} + \umal\rr \quad \forall i\in\leg.
	\end{equation}
	\begin{proof}
		From~\eqref{eq:deviation-split},
		it follows
		\begin{equation}
			\lim_{t\rightarrow\infty}\stateerr{i}{t} \le \lim_{t\rightarrow\infty}\statelegerr{i}{t} + \lim_{t\rightarrow\infty}\statemalerr{i}{t}
		\end{equation}
		and linearity of the expectation conditioned to $\tf<\infty$ yields
		\begin{equation} \label{eq:deviation_separated}
			\begin{aligned}
				\mean{\lim_{t\rightarrow\infty}\stateerr{i}{t}} &\le \mean{\lim_{t\rightarrow\infty}\statelegerr{i}{t}} + \mean{\lim_{t\rightarrow\infty}\statemalerr{i}{t}}\\
				&\stackrel{(i)}{\le}\eta(\uleg[]{} + \umal)
			\end{aligned}
		\end{equation}
		where $(i)$ uses \cref{lem:deviation-leg,lem:deviation-mal} and we omit the condition event in view of our convention discussed below~\eqref{eq:max-ss-deviation}.
		Applying the Markov inequality to~\eqref{eq:deviation_separated} readily yields~\eqref{eq:deviation-bound-prob}.
	\end{proof}
\end{thm}

The steady-state deviation caused by a specific choice of $\lam{t}$ can be assessed with the bound in~\eqref{eq:deviation-bound-prob},
which combines the monotonic behaviors of the two components $\uleg[]{}$ and $\umal$.
As a rule of thumb,
\revision{the deviation increases with $\gamma$ and $\tf$, and $\tf$ in turn decreases as $\meanleg^2$ and $\meanmal^2$ grow; see \cref{lem:misclassification-probability}.
	This establishes a quantitative link between detection time and decay rate of $\lam{t}$;
	the latter should offset misclassifications errors due to trust observations based, for instance, on their expectations.
	This design question is left to future research.}
	
On the other hand,
a small deviation may negatively affect the speed of~\eqref{eq:update-rule-regular} and fail multi-robot cooperation in the worst case.
We next quantify how the convergence speed of updates depends on the decay rate of $\lam{t}$.
            %!TEX ROOT = ../resilient_consensus_trust.tex

\subsection{Finite-Time Rate of Convergence}\label{sec:convergence-rate}

We aim to assess how fast the proposed resilient protocol converges to its steady-state.
Legitimate robots continuously inject inputs $\lam{t}\stateleg{0}$,
thus standard convergence tools for autonomous systems cannot be used.
This is possible in~\cite{Yemini22tro-resilienceConsensusTrust} even with malicious inputs under the assumption of bidirectional communication since,
after the classification time $\tf$,
legitimate robots follow a consensus protocol that is a reversible Markov chain.
However,
protocol~\eqref{eq:update-rule-regular} is not a Markov chain.
Results on convergence speed of the FJ model~\cite{Proskurnikov17ifacwc-timeVaryingFJ} and time-varying consensus~\cite{Blondel05cdc-consensusConvergence,Xiao06automatica-consensusMetropolisWeights} are inadequate to the present framework because they assume non-vanishing weights,
whereas $\lam{t}$ decays to zero in~\eqref{eq:update-rule-regular}.
Previous work~\cite{Ballotta24lcss-FJDiminishingCompetition} assumes doubly stochastic weights without malicious robots.

Next, we upper bound the expected convergence rate for the general case $\lam{t}\searrow0$.

\begin{prop}[Convergence speed of~\eqref{eq:update-rule-regular}]\label{thm:convergence-rate}
	Let \cref{lem:W-primitive,ass:trust-meaningful,ass:vanishing-lam,ass:initial-state-bound} hold and $\tf<\infty$ be fixed.
	Define the coefficients
	\begin{equation}\label{eq:D1}
		D_1 \doteq \max_{i\in\leg}\dfrac{|\mal\cap\neigh{i}|}{|\mal\cap\neigh{i}| + 1}, \quad \prodlamfin{s}{t} \doteq \prodlam{k}{s}{t}.
	\end{equation}
	Let $\specrad$ be the second largest eigenvalue modulus of $\Wlegtrue$,
	$m_\specrad+1\ge1$ the maximal size of  Jordan blocks associated with $\specrad$,
    and $m\ge1$ the maximal size of all Jordan blocks.
	It holds
	\begin{equation}\label{eq:conv-rate-bound}
		\norm[\infty]{\stateleg{t} - \statelegss} \le \eta\rho(t)
	\end{equation}
	where,
	for some $b>0$ which depends only on the (generalized) eigenvectors of $\Wlegtrue$,
	we define
	\begin{equation}
		\rho(t) \doteq \min\lb bm\sqrt{L}\rho_\leg(t) +
		2(\prodlamfin{\tf}{t-1}-\prodlamfin{\tf}{\infty})
		+ D_1\rho_\mal(t),2\rb
	\end{equation}
	\begin{multline}\label{eq:convergence-rate-bound-leg-t>tf}
		\rho_\leg(t) \doteq 
		\prodlamfin{0}{t-1}\binom{t-\tf+m_\specrad}{m_\specrad}\specrad^{t-\tf}
		\\
		+\sum_{k=0}^{t-1}\prodlamfin{k+1}{t-1}\lam{k}\binom{t-(\maxof{\tf}{(k+1)})+m_\specrad}{m_\specrad}\specrad^{t-(\maxof{\tf}{(k+1)})}
	\end{multline}
	\begin{multline}\label{eq:convergence-rate-bound-mal-t>tf}
		\rho_\mal(t) \doteq		\sum_{k=0}^{\tfmal-1} \prodlamfin{k}{t-1} \\ \cdot \lr bm\binom{t-\tfmal+m_\specrad}{m_\specrad}\specrad^{t-\tfmal} +  1-\prodlamfin{t}{\infty} \rr.
	\end{multline}
	\begin{proof}
		See \cref{ass:proof-convergence-rate}. 
	\end{proof}
\end{prop}

\begin{figure}
	\centering
	\includegraphics[width=.5\linewidth]{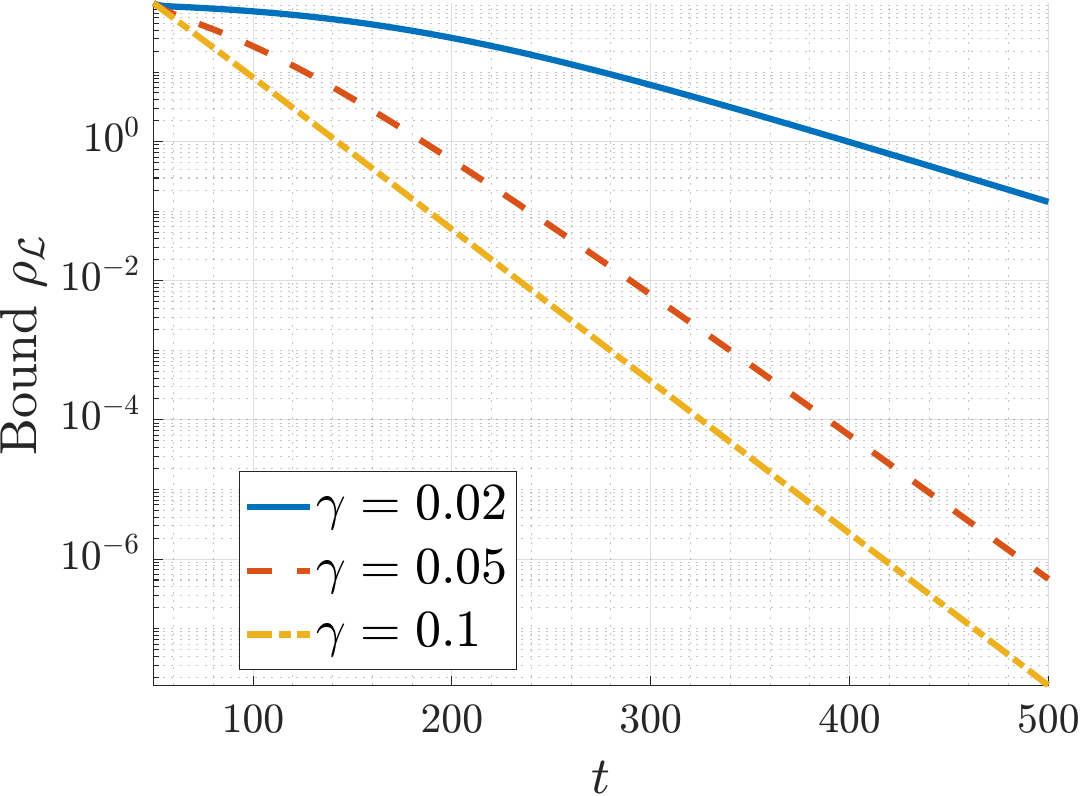}%
	\hfil
	\includegraphics[width=.5\linewidth]{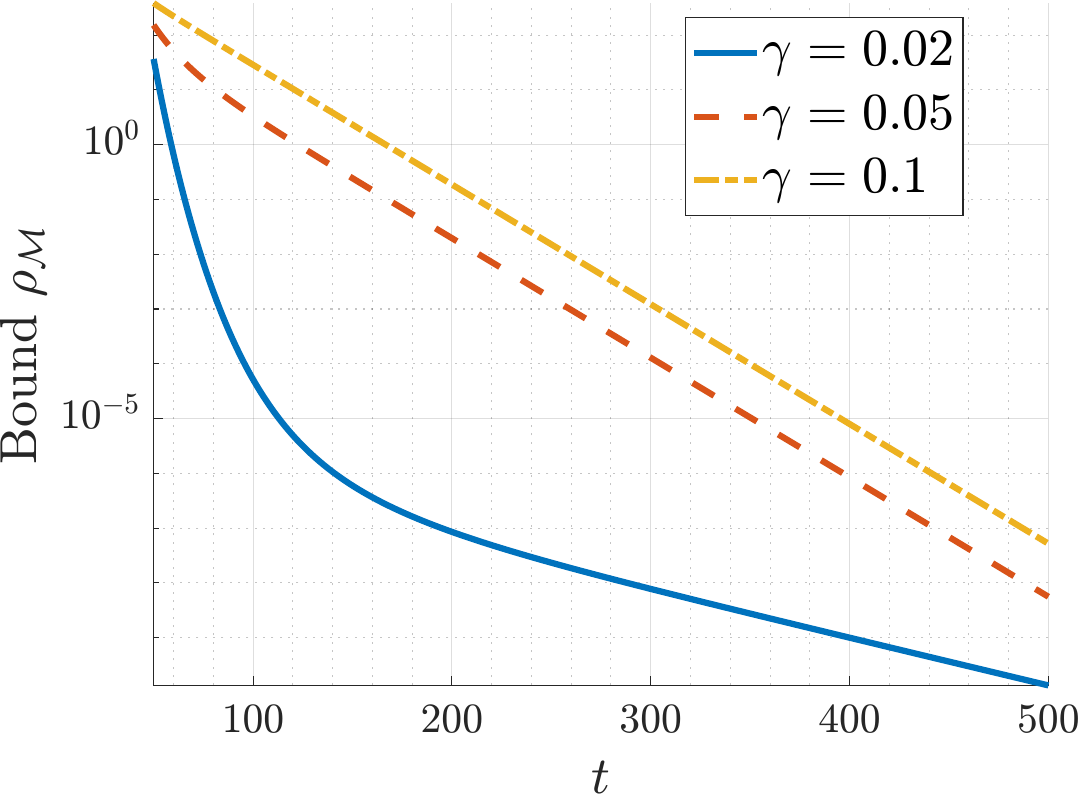}\\
	\includegraphics[width=.5\linewidth]{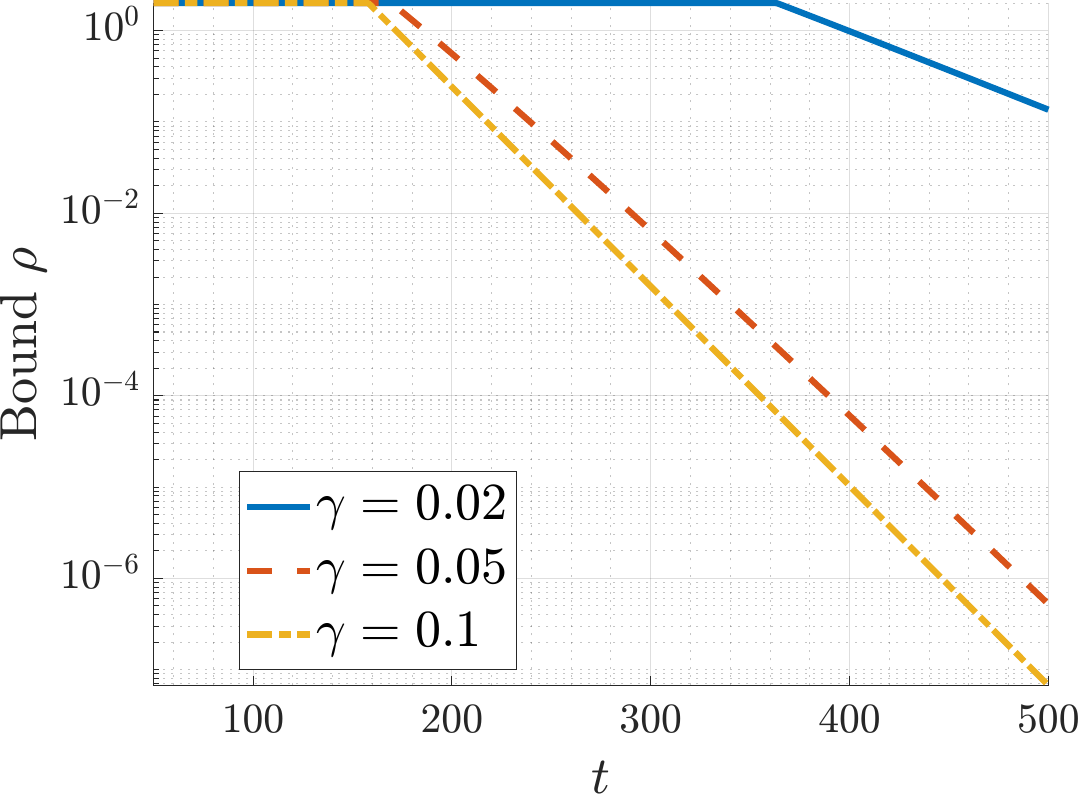}
	\caption{Upper bound $\rho(t)$ in \cref{thm:convergence-rate} on convergence rate for a random geometric graph with $L = 20$ and $\tf = \tfmal = 50$,
	and $\lam{t} = 0.9 \e^{-\gamma t}$.
	}
	\label{fig:rho}
\end{figure}

The bound $\rho(t)$ in~\eqref{eq:conv-rate-bound} is monotonically decreasing and vanishes as $t$ becomes large.
The terms $\rho_\leg(t)$ and $\rho_\mal(t)$ respectively bound the speed of convergence of the legitimate contribution $\statecontrleg{t}$ and of the malicious contribution $\statecontrmal{t}$.
From their expressions in~\eqref{eq:convergence-rate-bound-leg-t>tf} and~\eqref{eq:convergence-rate-bound-mal-t>tf},
we deduce that convergence of~\eqref{eq:update-rule-regular} is slower than geometric through the presence of factors $\prodlamfin{k-1}{t+1}$ and $\lam{k}$.
\Cref{fig:rho} illustrates the bound $\rho(t)$ and its two components for a random geometric graph.
Although $\rho_\leg(t)$ and $\rho_\mal(t)$ are initially loose,
due to the difficulty of addressing all couplings among agents,
they clearly suggest that the convergence rate is monotonic with the decay rate of $\lam{t}$.
In the next section we numerically show that the monotonic behavior hinted at by \autoref{fig:rho} is indeed observed on the actual convergence of~\eqref{eq:update-rule-regular}.
Although the finite sum in bound $\rho_\mal(t)$ associated with malicious inputs increases with $\gamma$,
the total bound $\rho(t)$ is mainly influenced by $\rho_\leg(t)$ and decreases with $\gamma$;
hence, smaller values of $\gamma$ slow down convergence in general.
This behavior is indeed expected because slowly decaying $\lam{t}$ (\ie small $\gamma$) forces legitimate robots to stick near the initial condition for a long time,
which overall slows down convergence.
This observation,
together with the discussion in \autoref{sec:deviation},
reveals a \emph{fundamental tradeoff between convergence speed and final deviation}.
Small deviation requires a slow decay of $\lam{t}$ to make updates resilient during the transient when most misclassifications occur;
fast convergence is achieved with a fast decaying $\lam{t}$.
\revision{From a practical perspective,
a designer should evaluate this tradeoff to tune the decay rate of $\lam{t}$, \eg the parameter $\gamma$ in~\eqref{eq:lambda},
based on contextual information such as estimates of $\tf$.}

\subsubsection*{Asymptotic regime with exponentially decaying $\lam{t}$} The following limits provide analytical insight on how the finite-time deviation bound~\eqref{eq:conv-rate-bound} in \cref{thm:convergence-rate} depends on the decay rate $\gamma$ of the parameter $\lam{t}$ in~\eqref{eq:lambda}.
For the sake of simplicity,
let $\lam{t} = \e^{-\gamma t}$ and assume $\tf\ge\tfmal\ge1$.
We address two regimes:
$\gamma\to\infty$,
where $\lam{t}\gtrapprox0$ and~\eqref{eq:update-rule-regular} practically reduces to standard consensus after few iterations;
$\gamma\to0$,
where $\lam{t}\lessapprox1$ and each legitimate robot sticks to its own initial condition for long time.
It holds
\begin{align}
    &\begin{multlined}\label{eq:convergence-rate-bound-lim-t>tf-gamma-infty}
        \lim_{\gamma\rightarrow\infty}\rho(t) = \min\left\lbrace bm\sqrt{L}\binom{t-\tf+m_\specrad}{m_\specrad}\specrad^{t-\tf}\right.\\
        + \left.bmD_1\tfmal \binom{t-\tfmal+m_\specrad}{m_\specrad}\specrad^{t-\tfmal},2\right\rbrace
    \end{multlined}\\
    &\lim_{\gamma\rightarrow0^+}\rho(t) = 2.\label{eq:convergence-rate-bound-lim-t>tf-gamma-0}
\end{align}
Limit~\eqref{eq:convergence-rate-bound-lim-t>tf-gamma-infty} reveals that,
when $\lam{t}$ decays fast,
after time $\tf$ protocol~\eqref{eq:update-rule-regular} reduces to the standard consensus with the true weights $\Wlegtrue$ and geometric convergence with (approximately) rate $\specrad$.
The factors $\sqrt{L}$ and $D_1\tfmal$ suggest that the new ``initial condition'' $\stateleg{\tf}$ for such a consensus protocol is far from the final consensus value because it is affected by misclassifications of respectively legitimate and malicious robots,
occurred before time $\tf$.
In particular,
$D_1$ expresses how many links connect legitimate to malicious robots,
hence the latter have many opportunities for attacks before being detected if $D_1$ or $\tfmal$ are large.
On the other hand,
limit~\eqref{eq:convergence-rate-bound-lim-t>tf-gamma-0} trivially shows that,
if $\lam{t}$ decays extremely slowly,
legitimate robots do not sensibly converge until a very long time.

From \cref{thm:convergence-rate}, we deduce an upper bound on the expected convergence rate after arbitrary finite iterations.
\begin{thm}[Expected convergence speed of~\eqref{eq:update-rule-regular}]
	Let $\rho(t;\tf,\tfmal)$ denote $\rho(t)$ in~\eqref{eq:conv-rate-bound} for given realizations of the classification times $\tf$ and $\tfmal$.
    Define
    \begin{gather}
        D_\leg \doteq \sum_{i\in\leg}|\leg\cap\neigh{i}|\label{eq:Dl}\\
        p(t) \doteq \min\lb D_\leg\frac{e^{-2t\meanleg^2}}{1-e^{-2\meanleg^2}} + D_\mal\frac{e^{-2t\meanmal^2}}{1-e^{-2\meanmal^2}},1\rb.\label{eq:bound-prob-tf}
    \end{gather}
	Under \cref{lem:W-primitive,ass:trust-meaningful,ass:vanishing-lam,ass:initial-state-bound},
	it holds
	\begin{equation}\label{eq:bound-conv-rate-exp}
		\mean{\norm[\infty]{\stateleg{t} - \statelegss}} \le \eta\min_{k\in[1,t]} \lr\rho(t;k,k) + 2p(k)\rr.
	\end{equation}
 \begin{proof}
    The bound $\rho(t;\tf,\tfmal)$ is increasing with $\tf$ and $\tfmal$.
    Thus, from the bound in~\eqref{eq:conv-rate-bound},
    it follows
    \begin{equation}\label{eq:finite-time-deviation-exp}
        \begin{aligned}
            \mean{\norm[\infty]{\stateleg{t} - \statelegss}}
            &\le \eta \mean{\rho(t;\tf,\tfmal)} \le \eta \mean{\rho(t;\tf,\tf)}\\
            &\begin{multlined}[t]
                \le \eta\min_{k \in [1,t]} \left(1\cdot\mean{\rho(t;\tf,\tf)|\tf\le k}\right. \\
                \left.+ \pr{\tf>k}\mean{\rho(t;\tf,\tf)|\tf> k}\right)
            \end{multlined}\\
            &\begin{multlined}[t]
                \le \eta\min_{k \in [1,t]} \left(\max_{s \in [1,k]}\rho(t;s,s)\right. \\
                +\left. \pr{\tf>k}\max_{s>k}\rho(t;s,s)\right).
            \end{multlined}
        \end{aligned}
    \end{equation}    
    Using~\cite[Lemma~2]{Yemini25tac-resilientDistributedOptim} with~\eqref{eq:bound-prob-tf}
    yields $\pr{\tf>k}\le p(k)$.
	This combined with~\eqref{eq:finite-time-deviation-exp} and $\rho(t;s,s)\le2$ in turn yield~\eqref{eq:bound-conv-rate-exp}.
	\end{proof}
\end{thm}
    %!TEX ROOT = ../resilient_consensus_trust.tex

\section{Experiments}\label{sec:simulations}

We test our resilient consensus algorithm motivated by vehicular platooning~\cite{Santini17-consensusBasedPlatooning}.
\revision{Since platoons are sparsely connected,
	they are susceptible to attacks~\cite{Taylor22-VehicularPlatoonCommunication} and unsuited to resilient methods that require dense network connectivity~\cite{LeBlanc13jsac-wmrs}.}

\revision{We consider two platoons with $\nicefrac{L}{2}=10$ legitimate vehicles each that merge in the presence of $M=10$ malicious vehicles.
	The vehicles are connected through a $2$-nearest neighbor topology, see \autoref{fig:topology};
	two legitimate vehicles, one from each platoon (\eg leading vehicles), share a link,
	while all other links connect vehicles within the same platoon.
	We arrange the malicious vehicles so that the nominal communication weight matrix satisfies \cref{lem:W-primitive}.
	This is the sparsest topology for resilient consensus:
	removing any legitimate link $(i,j)\in\leg\times\leg$ splits $\leg$ into two distinct connected components, directly invalidating \cref{lem:W-primitive}.
	
	To simulate the merging, the vehicles in each platoon initially travel at the same speed,
	which is different across the two platoons (see \autoref{fig:topology}),
	and all vehicles must agree on a common speed.
	Minimizing merging time requires vehicles to agree on their average initial speed.
	We consider positive speed values $\state{t}{i}\in[0,\eta]$ with $\eta=40$.
	Under \Cref{ass:initial-state-bound},
	the worst-case attack that maximizes the final deviation requires all malicious vehicles to collude and constantly communicate the value $\eta$ or $0$:
	this attracts all legitimate vehicles towards either extreme of the speed range and the farthest from nominal consensus.
	In our case, since the legitimate initial states are closer to $0$,
	we implement the worst-case colluding attack $\state{t}{i}\equiv\eta\,\forall i\in\mal$.}

We draw trust observations $\trust{i}{j}{t}$ from the distribution $\mathrm{Beta}(1.5,1)$ for legitimate and from $\mathrm{Beta}(0.75,1)$ for malicious vehicles.
These distributions satisfy \cref{ass:trust-meaningful} and induce nonnegligible chance of misleading trust observations: it holds $\pr{\trust{i}{j}{t}>0.5}\approx0.4$ for malicious vehicles $j\in\mal$ and $\pr{\trust{i}{j}{t}<0.5}\approx0.35$ for legitimate $j\in\leg$.
This significant portion of ``wrong'' trust observations makes misclassification probability decay slowly,
proving that~\eqref{eq:update-rule-regular} enhances resilience even with many misclassifications.
\revision{In the experiments, we average all runs over $1000$ Monte Carlo iterations.}

\begin{figure}
	\centering
	\includegraphics[width=\linewidth]{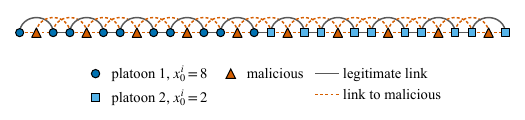}
	\caption{\revision{Communication topology. Every legitimate vehicle has two legitimate neighbors and two malicious neighbors.}
	}
	\label{fig:topology}
\end{figure}

\begin{figure}
	\centering
	\includegraphics[width=\linewidth]{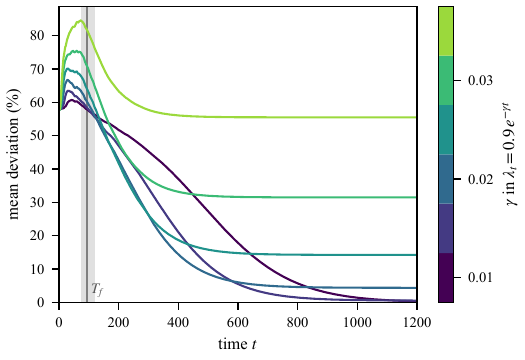}
	\caption{Relative mean deviation from nominal consensus of legitimate vehicles.
		Parameter $\lam{t}$ decays exponentially fast with rate $\gamma$ given by the colorbar.
		\revision{Vertical line and band: median and interquartile range of $\tf$.}}
	\label{fig:multiple_lambdas}
\end{figure}

\begin{figure}
	\centering
	\includegraphics[width=.8\linewidth]{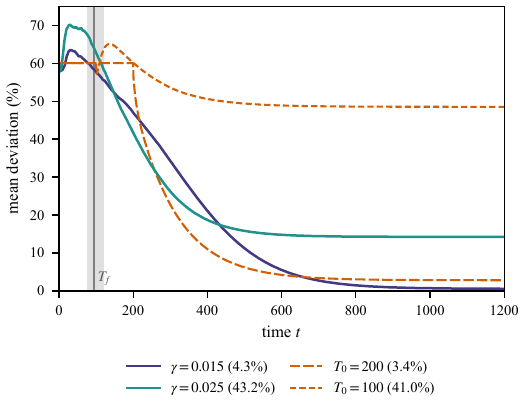}
	\caption{Mean distance from nominal consensus of legitimate vehicles.
		\revision{Solid: our method with $\lam{t}=0.9\e^{-\gamma t}$; other curves:} the algorithm in~\cite{Yemini22tro-resilienceConsensusTrust} with \revision{different $T_0$. In parentheses: share of runs ending farther than $0.1$ from nominal consensus. Vertical line and band as in \Cref{fig:multiple_lambdas}.}}
	\label{fig:vst0}
\end{figure}

We plot the relative mean deviation of legitimate vehicles in \Cref{fig:multiple_lambdas},
which shows that~\eqref{eq:update-rule-regular} provides fine granularity to trade fast convergence for small deviation, in agreement with the theory.
A designer can smoothly control this tradeoff by setting $\gamma$ or, more generally, the decay rate of $\lam{t}$.

Next, we compare~\eqref{eq:update-rule-regular} against the algorithm in~\cite{Yemini22tro-resilienceConsensusTrust},
which requires the vehicles to accrue trust observations for $T_0$ time-steps before starting consensus.
While one would ideally set $T_0 = \tf$ to discard all malicious messages,
setting $T_0$ is practically difficult as $\tf$ is unknown:
a short $T_0$ means that legitimate vehicles trust many malicious messages; a large $T_0$ slows down convergence.

\revision{In \autoref{fig:vst0},
	we plot the deviation with two choices for $\lam{t}$ and with $T_0\in\{100,200\}$, averaging over $1000$ Monte Carlo iterations.
	In particular,
	after a grid search for $\gamma$,
	we pair each $T_0$ with the decay rate $\gamma$ that yields the closest failure rate (\ie fraction of failed runs),
	where we define a run successful if its deviation after $T=1200$ time-steps is less than $0.1$.
	When $T_0 < \tf$, the method in~\cite{Yemini22tro-resilienceConsensusTrust} almost always fails,
	proving it not robust.
	For $T_0 = 200$, its failing runs feature a mean final deviation of $4.1$ against $0.58$ for our method with $\gamma = 0.015$,
	which also yields a smaller deviation across all runs ($0.027$ vs. $0.14$).
	Overall,
	at similar failure rates,
	our method reaches closer to nominal consensus.
	Further,
	\autoref{fig:vst0} suggests that it may provide an advantage in the short run by reducing the deviation at time $T_0$,
	see the height of the curve with $\gamma=0.015$ at time $T_0=200$.}

\revision{Finally, we compare exponential decay with polynomial decay $\lam{t} = c(1+\nicefrac{t}{\tau})^{-p}$ for several degrees $p$ and time scales $\tau$, keeping $c = 0.9$.
	\Cref{fig:lambda_designs} reports the share of failed runs (defined as above) versus median time to empirical agreement $T_\textnormal{a}=\min\{t:\max_{i,j\in\leg}|\state{s}{i}-\state{s}{j}|\le0.1 \,\forall s\ge t\}$.
	Markers on polynomial-decay curves represent each of the tested $16$ time scales for $\tau$.
	All designs exhibit the speed-deviation tradeoff:
	as $\lam{t}$ decays more slowly, \ie smaller $\gamma$ and $p$ and larger $\tau$, fewer runs are strongly affected by attacks, but agreement takes longer.
	Exponential decay exhibits the best tradeoff, approaching the bottom-left corner closer than all polynomial decays.
	In fact, we observe a clear monotonic behavior as the degree of polynomial $\lam{t}$ decreases, with largest degrees yielding better Pareto frontiers.}

\begin{figure}
	\centering
	\includegraphics[width=\linewidth]{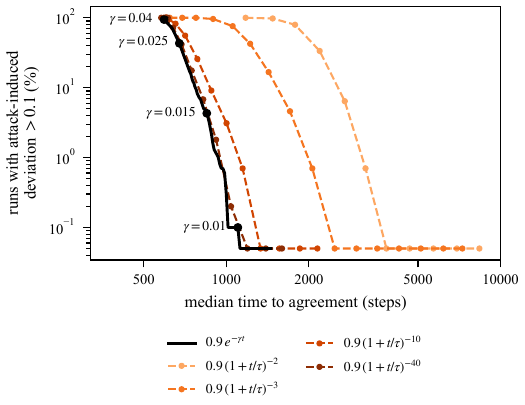}
	\caption{\revision{Comparison between exponential and polynomial decay of $\lam{t}$.
	Markers at $0.05\%$ mean no runs exhibit deviation beyond $0.1$.}}
	\label{fig:lambda_designs}
\end{figure}
    %!TEX ROOT = ../resilient_consensus_trust.tex

\section{Current Limitations and Outlook}\label{sec:limitations}

While this manuscript establishes fundamental convergence properties of protocol~\eqref{eq:update-rule-regular} and the speed-deviation tradeoff it entails,
its analysis and design require effort to work towards practical deployment.
One critical aspect involves the design of $\lam{t}$ that is currently assumed to be performed once offline in a centralized fashion.
Even though \cref{prop:consensus} establishes convergence to a consensus under any vanishing $\lam{t}$, its choice affects deviation from nominal and speed of convergence, making it practically relevant.
Since precise knowledge of statistics $\meanleg$, $\meanmal$, and $\tf$ is not realistic,
legitimate robots would ideally tune $\lam{t}$ during operations in a decentralized fashion.
However, this notably complicates analytical study and establishing performance bounds: our previous work~\cite{Ballotta24lcss-FJDiminishingCompetition} proves that the nominal consensus is \emph{not} recovered with local parameters $\lam{t}^i$ even without adversaries, which prevents us from adopting naive adaptive rules that tune $\lam{t}^i$ from each robot's local information.
Moreover,
adapting $\lam{t}^i$ based on data exchange with neighbors in $\neigh{i}$ or $\neigh[t]{i}$ introduces a new coupling between physics-based classification and data-based distributed protocol, whose decoupling is the key strength of physical trust-based approaches.
Future work on local and/or online design must therefore balance performance in adversary-free conditions against robustness to unknown malicious robots without compromising convergence guarantees obtained from physical trust information.

A second relevant aspect is the robustness of~\eqref{eq:update-rule-regular} to nonideal wireless communication.
Quantization effects can be approximated with additive noise and handled by resorting to mean-square or robust analysis~\cite{6851917}.
However, explicitly addressing quantization-dependent design or codesign of distributed protocol and encoding-decoding scheme has proved effective in recovering the nominal solution in consensus and distributed optimization~\cite{5286289,9937075}.
On the other hand, desynchronization, delays, time-varying links, and packet loss require dedicated design to ensure convergence, making the extension of our method to such nonidealities nontrivial~\cite{4738941}.
On a positive note, as stated in \cref{rem:trust-non-iid}, the characterization of trust variables in \cref{def:alpha} can be relaxed, making our methodology fundamentally valid even if trust observations are correlated across time or links.
    %!TEX ROOT = ../resilient_consensus_trust.tex

\section{Conclusions}\label{sec:conclusions}

We have proposed a resilient consensus algorithm that considers the confidence about trust observations from  a physical channel.
Robots scale messages from trusted neighbors by $1-\lam{t}$,
where $\lam{t}$ vanishes over time,
making the protocol resilient to misclassification.
Our algorithm induces a tradeoff between deviation from nominal consensus and convergence speed that can be controlled through the decay rate of $\lam{t}$.

Opportunities for future research are multifold.
Practical design of the confidence parameter $\lam{t}$ should be studied,
possibly using trust statistics for offline design and/or real-time observations $\trust{i}{j}{t}$ for adaptation online.
Decentralized strategies to locally tune $\lam{t}$ at each robot should be investigated to eliminate the need for a centralized design.
\revision{Extension to nonideal network, e.g., with delays, is relevant for deployment.}

    \section*{Acknowledgment}

M.~Yemini thanks Prof. Reuven Cohen for an enriching discussion regarding the finer points of the convergence of random variables and conditional expectations.  

    \appendices
    \numberwithin{equation}{section}
    \crefalias{section}{appendix}
    \crefalias{subsection}{appendix}
    \crefalias{subsubsection}{appendix}
    
    %!TEX ROOT = ../resilient_consensus_trust.tex

\section{Proof of \cref{prop:consensus}}
\label{app:consensus}

Let $\maxof{a}{b} \doteq \max\{a,b\}$ and define the two products
\begin{equation}\label{eq:def-prod-infty-lam}
	\prodinflam{k_0} \doteq \prodlam{k}{\maxof{k_0}{\tf}}{\infty}, \text{ and } \prodlamW{k_0} \doteq \prodW{k}{k_0}{\tf-1}.
\end{equation}
From \cref{lem:final-misclassification,lem:W-primitive},
there exists a finite time $\tf\ge0$ such that,
almost surely,
for every $k_0\ge0$,
\begin{equation}\label{eq:prod-leg-influence-infty-1}
	\begin{aligned}
		\prodW{k}{k_0}{\infty} 	&= \prodlam{k}{\maxof{k_0}{\tf}}{\infty}\Wlegtrue\prodW{k}{k_0}{\tf-1}  \\
												&= \one v^\top \prodinflam{k_0} \prodlamW{k_0}
	\end{aligned}
\end{equation}
The second product in~\eqref{eq:def-prod-infty-lam} is empty if $k_0\ge\tf$,
\ie $\prodlamW{k_0} = I$ and the matrix product in~\eqref{eq:prod-leg-influence-infty-1} is simply a scaled version of $\one v^\top$.
In view of~\eqref{eq:leg-state-contr-leg}--\eqref{eq:leg-state-contr-mal},
we separately consider the two contributions by legitimate and malicious robots.
If we can prove that each contribution achieves a consensus,
then the claim~\eqref{eq:consensus} trivially follows from~\eqref{eq:dynamics} and properties of the limit.

%!TEX ROOT = ../resilient_consensus_trust.tex

\subsubsection*{Contribution by legitimate robots}

From the definition~\eqref{eq:leg-state-contr-leg} and~\eqref{eq:prod-leg-influence-infty-1},
almost surely it holds
\begin{equation}\label{eq:leg-agents-influence-convergence}
	\begin{aligned}
		\lim_{t\rightarrow\infty}\statecontrleg{t}	& = \one v^\top \prodinflam{0} \prodlamW{0}\stateleg{0} + \sum_{k=0}^\infty \lr\Wlegtrue\rr^{t-k} \prodinflam{k+1} \prodlamW{k+1}\lam{k} \stateleg{0}\\
		& = \begin{multlined}[t]
			\one v^\top \lr\prodinflam{0}\prodlamW{0} \stateleg{0} + \sum_{k=0}^{\tf-2} \prodinflam{k+1}\prodlamW{k+1}\lam{k}\stateleg{0}\rr\\
			+ \sum_{k=\tf-1}^\infty \lr\Wlegtrue\rr^{t-k} \prodinflam{k+1} \lam{k} \stateleg{0}
		\end{multlined}\\
		& \stackrel{(i)}{=} \begin{multlined}[t]
			\one v^\top \lr\prodinflam{0}\prodlamW{0} + \sum_{k=0}^{\tf-2} \prodinflam{k+1}\prodlamW{k+1}\lam{k}\rr \stateleg{0}\\
			+ \one v^\top \sum_{k=\tf-1}^\infty \prodinflam{k+1} \lam{k} \stateleg{0}.
		\end{multlined}
	\end{aligned}
\end{equation}
Equality $(i)$ follows from the convergence result in~\cite{Ballotta24lcss-FJDiminishingCompetition} for the FJ dynamics with vanishing $\lam{k}$\arxiv{,
by which we also get that the sum of the series in the last line of~\eqref{eq:leg-agents-influence-convergence} is well defined and is nonzero if and only if $\lam{k}$ is summable~\cite{Trench99amm-infiniteProducts}}.
The vector in the last line is well defined and correspond to a consensus.
%!TEX ROOT = ../resilient_consensus_trust.tex

\subsubsection*{Contribution by malicious robots}
\label{app:consensus-mal}

From the definition~\eqref{eq:leg-state-contr-mal} and~\eqref{eq:prod-leg-influence-infty-1},
almost surely it holds $\tfmal<\infty$ and
\begin{equation}\label{eq:mal-agents-influence-convergence}
	\begin{aligned}
		\lim_{t\rightarrow\infty}\statecontrmal{t}	&=\sum_{k=0}^\infty \one v^\top \prodinflam{k+1} \prodlamW{k+1} (1-\lam{k})\Wmal{k}\statemal{k}\\
		&= \one v^\top\underbrace{\sum_{k=0}^{\tfmal-1} \prodinflam{k+1} \prodlamW{k+1} (1-\lam{k})\Wmal{k}\statemal{k}}_{\doteq\statemaltil{}}.
	\end{aligned}
\end{equation}

Combining~\eqref{eq:dynamics} with~\eqref{eq:leg-agents-influence-convergence}--\eqref{eq:mal-agents-influence-convergence},
we conclude that,
almost surely,
$\lim_{t\rightarrow\infty}\stateleg{t} = \one v^\top(\statelegtil{}+\statemaltil{})$ with $\statelegtil{}$ given by~\eqref{eq:leg-agents-influence-convergence}.
Thus, the claim~\eqref{eq:consensus} holds with $\statelegss = v^\top(\statelegtil{}+\statemaltil{})$. \qedsymbol
    %!TEX ROOT = ../resilient_consensus_trust.tex

\section{Proof of \cref{lem:deviation-leg}}
\label{app:proof-deviation-leg}

Before proceeding with the deviation bound for the legitimate contribution,
we state an ancillary result used in the proof.

\begin{cor}[Difference of sub-stochastic matrices~\protect{\cite[Lemma~4]{Yemini22tro-resilienceConsensusTrust}}]\label{lem:matrix-difference}
	Let $\ell > 0$ and $X,Y\in\Real{N\times N}$ be two sub-stochastic matrices such that $[X]_{ii} \ge \ell$ and $[Y]_{ii} \ge \ell$ for $i = 1,\dots,N$.
	Then,
	$\max_i[|X-Y|\one]_{i} \le 2(1-\ell)$.
\end{cor}

We next bound the two expectations in~\eqref{eq:leg-err-markov-limit}.
	       %!TEX ROOT = ../resilient_consensus_trust.tex

\subsubsection*{Deviation caused by legitimate autonomous dynamics}

Let $T(t)\le t$ be the first instant such that the true weights are consistently recovered through time $t-1$:
\begin{equation}\label{eq:T(t)}
	T(t) \doteq \min \lb k\ge0 : \Wleg{s} = \Wlegtrue, s = k,\dots,t-1 \rb,
\end{equation}
where we define $\min\{\emptyset\}\doteq t$.
Time $T(t)$ is a random variable that is nondecreasing in $t$.
By \cref{lem:final-misclassification},
there exists $\tf\in\Realp{}$ such that $T(t) \le \tf$ for all $t\in\Realp{}$ almost surely.
Define
\begin{equation}\label{eq:delta-wtilde1}
	\Delta\Wlegerr[\text{aut}]{t}\doteq \prodlam{k}{0}{t-1}\left(\prod_{k=0}^{T(t)-1}W_k^{\mathcal{L}}-\prod_{k=0}^{T(t)-1}\overline{W}^{\mathcal{L}}\right)
\end{equation}
and its limit $\Delta\Wlegerr[\text{aut}]{\infty}\doteq\lim_{t\rightarrow\infty} \Delta\Wlegerr[\text{aut}]{t}$ evaluates
\begin{equation}\label{eq:delta-wtilde1-ss}
	\Delta\Wlegerr[\text{aut}]{\infty} =\prodlam{k}{0}{\infty}\lr\prod_{k=0}^{\tf-1}W_k^{\mathcal{L}}-\prod_{k=0}^{\tf-1}\overline{W}^{\mathcal{L}}\rr.
\end{equation}
From \cref{lem:W-primitive},
it follows that,
if $\tf<\infty$,
then
\begin{equation}\label{eq:delta-w1-ss-bound}
	\begin{aligned}
		\lim_{t\rightarrow\infty}\left|\ls\Wlegerr[\text{aut}]{t}\stateleg{0}\rs_i\right| &= \left|\ls \lr\prodWtrue{k}{\tf}{\infty}\rr\Delta\Wlegerr[\text{aut}]{\infty}\stateleg{0}\rs_i\right|\\
		&\stackrel{(i)}{\le} \max_{i\in\leg} \left|\ls \Delta\Wlegerr[\text{aut}]{\infty}\stateleg{0}\rs_i\right|\\
		&\stackrel{(ii)}{\le} \eta \max_{i\in\leg}\ls\left|\Delta\Wlegerr[\text{aut}]{\infty}\right|\one\rs_i\\
		&\stackrel{(iii)}{\le}2\eta\prodlam{k}{0}{\infty}\lr1-\dfrac{1}{\lr\dmax+1\rr^{\tf}}\rr
	\end{aligned}
\end{equation}
where $(i)$ because $\Wlegtrue$ is stochastic,
$(ii)$ from~\cref{ass:initial-state-bound},
and $(iii)$ from~\cref{lem:matrix-difference} in view of~\eqref{eq:delta-wtilde1-ss} and (see~\eqref{eq:weights-true} and~\eqref{eq:weights-trust})
\begin{equation}\label{eq:bound-wii}
	\ls\Wleg{t}\rs_{ii} \ge \dfrac{1}{\dmax+1}, \qquad \ls\Wlegtrue\rs_{ii} \ge \dfrac{1}{\dmax+1}.
\end{equation}
Next,
we find an upper bound to the infinite product in~\eqref{eq:delta-w1-ss-bound}.
\arxiv{The bound~\eqref{eq:delta-w1-ss-bound} is increasing with $\gamma$.
We next develop an upper bound that preserves this behavior consistently.}
It holds:
\begin{equation}\label{eq:bound-series}
	\begin{aligned}
		\prod_{k=0}^{\infty}\left(1-\lambda_k\right)
		\journalVersion{}{&= \exp\left(\sum_{k=0}^{\infty}\ln\left(1-ce^{-\gamma k}\right)\right)\\
			&}\leq \exp\left(\int_{0}^{\infty}\ln\left(1-ce^{-\gamma (k+1)}\right)\diff k\right).
	\end{aligned}
\end{equation}
The following equality holds from the definition of the dilogarithm function $\text{Li}_2$ and a change of variable:
\begin{equation}
	\int_{0}^{\infty}\ln\left(1-ce^{-\gamma (k+1)}\right)\diff k = -\frac{\text{Li}_2\lr ce^{-\gamma}\rr}{\gamma}.
\end{equation}
Let
\begin{equation}\label{eq:sbar}
	s(x)\doteq\frac{x - x\ln(1 - x) + \ln(1 - x)}{x}.
\end{equation}
For $|x|\leq 1$,
recall the identities $s(x)=\sum_{k=1}^{\infty}\frac{x^k}{k(k+1)}$ and $\text{Li}_2(x)=\sum_{k=1}^{\infty}\frac{x^k}{k^2}$.
Let $z(\gamma;k)\doteq-\frac{1}{\gamma}s(ce^{-\gamma(k+1)})$.
It follows
\begin{equation}\label{eq:inf-prod-upper-bound}
		-\frac{\text{Li}_2\left(ce^{-\gamma}\right)}{\gamma} 
		\le -\frac{1}{\gamma}\sum_{k=1}^{\infty}\frac{\lr ce^{-\gamma}\rr^k}{k(k+1)}=z(\gamma;0).
\end{equation}
Finally,
from~\eqref{eq:delta-w1-ss-bound},~\eqref{eq:bound-series}, and~\eqref{eq:inf-prod-upper-bound},
and assuming $\tf<\infty$,
the first expectation in~\eqref{eq:leg-err-markov-limit} can be upper bounded as follows,
\begin{equation}\label{eq:w1-ss-bound}
	\begin{aligned}
        \mean{\lim_{t\rightarrow\infty}\left|\ls\Wlegerr[\text{aut}]{t}\stateleg{0}\rs_i\right|}	&\le 2\eta\e^{z(\gamma;0)} \mean{1-\dfrac{1}{\lr\dmax+1\rr^{\tf}}} \\
        &\stackrel{(i)}{\le} 2\eta\e^{z(\gamma;0)} \lr1-\dfrac{1}{\lr\dmax+1\rr^{\mean{\tf}}}\rr
	\end{aligned}
\end{equation}
where $(i)$ follows from Jensen's inequality.
This proves~\eqref{eq:deviation-leg-one-bound}.
	       %!TEX ROOT = ../resilient_consensus_trust.tex

\subsubsection*{Deviation caused by legitimate input}

We proceed in the same spirit of the derivation in the previous section.
From~\eqref{eq:Winleg} and~\eqref{eq:T(t)},
we rewrite $\Winleg{t}$ as
\begin{multline}
    \Winleg{t}= \sum_{k=0}^{T(t)-2} \lr\prodlam{s}{k+1}{t-1}\rr\lr\prodWleg{s}{k+1}{t-1}\rr\lam{k}\\
    + \sum_{k=\maxof{(T(t)-1)}{0}}^{t-1} 
        \lr\prodWss{s}{k+1}{t-1}\rr\lam{k}.
\end{multline}
%Note that $\Winleg{t}=0$ if $T(t)\le1$.
For $\tf<\infty$,
its limit is
\begin{multline}\label{eq:lim-Win-split}
	\Winleg{\infty} = 
	\sum_{k=0}^{\tf-2} \lr\prodlam{s}{k+1}{\infty}\rr\lr\prodWleg{s}{k+1}{\infty}\rr\lam{k}\\
	+ \sum_{k=\maxof{(\tf-1)}{0}}^{\infty} \lr\prodWss{s}{k+1}{\infty}\rr\lam{k}.
\end{multline}
The infinite summation in~\eqref{eq:lim-Win-split} is a tail of the infinite summation associated with true weights in~\eqref{eq:leg-err-contribution-input},
and thus these two cancel out.
In analogy to~\eqref{eq:delta-wtilde1-ss},
define
\begin{equation}\label{eq:input-deviation-k}
	\Delta\Wlegerr[\text{in}]{k} \doteq \lr\prodlam{s}{k+1}{\infty}\rr\lr\prodWleg{s}{k+1}{\tf-1} -\prodWtrue{s}{k+1}{\tf-1}\rr.
\end{equation}
\arxiv{Note that $\Delta\Wlegerr[\text{in}]{k}=0$ if $\tf<k+1$.}
Applying the triangle inequality,
properties of sub-stochastic matrices,
and \cref{ass:initial-state-bound} analogous to~\eqref{eq:delta-w1-ss-bound} yields
\begin{equation}\label{eq:input-deviation-diff}
	\begin{aligned}
		\lim_{t\rightarrow\infty}\left|\ls\Wlegerr[\text{in}]{t}\stateleg{0}\rs_i\right|
		&= \left|\ls\sum_{k=0}^{\tf-2} \lam{k}\lr\prodWtrue{s}{\tf}{\infty}\rr\Delta\Wlegerr[\text{in}]{k}\stateleg{0}\rs_i\right|\\
		&\le\sum_{k=0}^{\tf-2} \lam{k} \left|\ls\lr\prodWtrue{s}{\tf}{\infty}\rr\Delta\Wlegerr[\text{in}]{k}\stateleg{0}\rs_i\right|\\
		&\le\sum_{k=0}^{\tf-2} \lam{k} \max_{i\in\leg} \left|\ls \Delta\Wlegerr[\text{in}]{k}\stateleg{0} \rs_i\right|\\
		&\le\sum_{k=0}^{\tf-2} \lam{k} \eta\max_{i\in\leg} \ls \left| \Delta\Wlegerr[\text{in}]{k}\right|\one\rs_i
	\end{aligned}
\end{equation}
and,
from~\eqref{eq:input-deviation-k} and~\eqref{eq:bound-wii},
it follows
\begin{multline}\label{eq:lim-wtilde2in}
	\lim_{t\rightarrow\infty} \left|\ls\Wlegerr[\text{in}]{t}\stateleg{0}\rs_i\right| \le \\
	2\eta \sum_{k=0}^{\tf-2} \lr\prodlam{s}{k+1}{\infty}\rr\lam{k}\lr1-\dfrac{1}{\lr\dmax+1\rr^{\tf-k-1}}\rr.
\end{multline}
The products in~\eqref{eq:lim-wtilde2in} can be bounded akin to~\eqref{eq:bound-series}--\eqref{eq:inf-prod-upper-bound} as
\begin{equation}\label{eq:bound-z}
	\prodlam{s}{k+1}{\infty} < \e^{z(\gamma;k+1)}.
\end{equation}
Subbing~\eqref{eq:bound-z} into~\eqref{eq:lim-wtilde2in} and taking expectation yields~\eqref{eq:deviation-leg-two-bound}.
	%!TEX root = ../resilient_consensus_trust.tex

\section{Proof of \cref{lem:deviation-mal}}
\label{app:proof-deviation-mal}

From~\eqref{eq:leg-state-contr-mal} and~\eqref{eq:state-mal-error},
it follows
\begin{equation}\label{eq:state-mal-error-bound}
	\begin{aligned}
		\statemalerr{i}{t} &= \left|\ls \sum_{k=0}^{t-1} \Winmal{k}{t}\statemal{k}\rs_i\right|\\
		&\stackrel{(i)}{\le} \sum_{k=0}^{t-1} \left|\ls  \Winmal{k}{t}\statemal{k}\rs_i\right|\stackrel{(ii)}{\le} \eta \sum_{k=0}^{t-1} \ls \Winmal{k}{t}\one\rs_i\\
		&\stackrel{(iii)}{\le} 
        \begin{multlined}[t]
            \eta \sum_{k=0}^{t-2} (1-\lam{k+1}) (1-\lam{k})\max_{i\in\mathcal{L}}\ls \Wmal{k}\one\rs_i \\ + (1-\lam{t-1})\max_{i\in\mathcal{L}}\ls \Wmal{t-1}\one\rs_i
        \end{multlined}
	\end{aligned}
\end{equation}
where 
$(i)$ is the triangle inequality,
$(ii)$ follows from \cref{ass:initial-state-bound},
and $(iii)$ because $\Winmal{k}{t}$ are sub-stochastic matrices and $\lb\lam{t}\rb_{t\ge0}$ is a decreasing sequence featuring $0 < 1-\lam{t} < 1$.
The weights given to malicious robots are upper bounded as
\begin{equation}\label{eq:weights-mal-upper-bound}
	\ls \Wmal{t}\one\rs_i = \sum_{j=1}^M \ls \Wmal{t}\rs_{ij} \le \sum_{j\in\mal}\dfrac{1}{2}\one[\beta_{ij}(t)\ge0]
\end{equation}
and further
\begin{equation}
	\max_{i\in\mathcal{L}}\ls \Wmal{t}\one\rs_i \le \sum_{i\in\leg}\sum_{j\in\mal}\dfrac{1}{2}\one[\beta_{ij}(t)\ge0].
\end{equation}
It follows
\begin{equation}\label{eq:weight-mal-upper-bound-max}
	\begin{aligned}
		\mean{\max_{i\in\mathcal{L}}\ls\Wmal{t}\one\rs_i} &\le  \mean{\sum_{i\in\leg}\sum_{j\in\mal}\dfrac{1}{2}\one[\beta_{ij}(t)\ge0]}\\
		&=\sum_{i\in\leg}\sum_{j\in\mal}\dfrac{1}{2}\pr{\beta_{ij}(t)\ge0} \\
		&\le\dfrac{1}{2}\sum_{i\in\leg}\sum_{j\in\mal\cap\neigh{i}}\e^{-2(t+1)\meanmal^2}\\
		&=\dfrac{D_\mal}{2}\e^{-2(t+1)\meanmal^2}.
	\end{aligned}
\end{equation}
Let us denote by $\tfmal[t]$ the first time-step when all malicious robots are correctly classified throughout the time interval $\{\tfmal[t],\dots,t-1\}$ with $\tfmal[t]\doteq t$ if misclassifications occur at time $t-1$.
Then,
it follows that $\Wmal{k}=0$ for $\tfmal[t]\le k< t$.
Combining~\eqref{eq:state-mal-error-bound} and~\eqref{eq:weight-mal-upper-bound-max} and assuming $\tfmal[t]<t-1$, which is verified at the limit $t\to\infty$ if $\tfmal<\infty$, yields
\begin{equation}\label{eq:state-mal-error-mean-bound}
	\begin{aligned}
		\mean{\statemalerr{i}{t}}
        &\le
        \begin{aligned}[t]
            &\mean{\eta \sum_{k=0}^{t-2} (1-\lam{k+1}) (1-\lam{k}) \max_{i\in\mathcal{L}}\ls \Wmal{k}\one\rs_i}\\
            &+ \mean{(1-\lam{t-1})\max_{i\in\mathcal{L}}\ls \Wmal{t-1}\one\rs_i}
        \end{aligned}\\
			&= \eta \sum_{k=0}^{\tfmal[t]-1} (1-\lam{k+1}) (1-\lam{k})\mean{\max_{i\in\mathcal{L}}\ls \Wmal{k}\one\rs_i}\\
			&\le \dfrac{D_\mal\eta}{2} \xi(\tfmal[t])
	\end{aligned}
\end{equation}
where we define
\begin{equation}
	\xi(\tfmal[t]) \doteq \sum_{k=0}^{\tfmal[t]-1} (1-\lam{k+1}) (1-\lam{k}) \e^{-2(k+1)\meanmal^2}.
\end{equation}
By \cref{lem:final-misclassification},
it holds $\tfmal[t] \le \tfmal$ for all $t\ge0$.
Also,
$\tfmal[t]$ is nondecreasing.
It follows that,
if $\tfmal<\infty$,
then $\lim_{t\rightarrow\infty}\xi(\tfmal[t]) = \xi(\tfmal)$ can be explicitly computed as
\begin{multline}\label{eq:xi}
    \mathllap{\xi}(\tfmal) = \dfrac{1 - \e^{-2\meanmal^2\tfmal}}{\e^{2\meanmal^2}-1} 
    - \dfrac{c(1+\e^{-\gamma})\lr1-\e^{-(\gamma+2\meanmal^2)\tfmal}\rr}{\e^{2\meanmal^2}-\e^{-\gamma}} \\
    + \dfrac{c^2\e^{-\gamma}\lr1-\e^{-2(\gamma+\meanmal^2)\tfmal}\rr}{\e^{2\meanmal^2}-\e^{-2\gamma}}.
\end{multline}
Under the condition $\tfmal<\infty$,
the limit $\lim_{t\rightarrow\infty}\xi(\tfmal[t])$ and the expectation in~\eqref{eq:state-mal-error-mean-bound} can be exchanged because the limit yields a finite sum.
It follows that
\begin{equation} \label{eq:bound_first_half}
    \mean{\lim_{t\rightarrow\infty}\statemalerr{i}{t} \;|\; \tfmal<\infty}
    \le\dfrac{D_\mal\eta}{2}\;\mean{\xi(\tfmal)}.
\end{equation}
We now compute an upper bound for $\mean{\xi(\tfmal)}$.
By definition,
\begin{equation}\label{eq:xi-mean-def}
	\mean{\xi(\tfmal)}=\sum_{k=0}^{\infty}\xi(k)\pr{\tfmal=k}.
\end{equation}
The probability of final misclassification time can be bounded as $\pr{\tfmal=k} \le D_\mal\e^{-2\meanmal^2 k}$,
see \cref{app:final-time-misclassification-all}.
\arxiv{Note that this bound is conservative because,
although it holds $\tfmal<\infty$ with probability $1$,
it is not possible to identify a constant $T_{\text{max}}\in\mathbb{N}$ \emph{a priori} such that $\pr{\tfmal>T_{\text{max}}}=0$. }%
Combining this with~\eqref{eq:xi-mean-def} yields
\begin{equation}\label{eq:bound_second_half}
		\mean{\xi(\tfmal)}	\le\sum_{k=0}^{\infty}\xi(k)D_\mal\e^{-2\meanmal^2 k} \stackrel{\eqref{eq:xi}}{=} D_\mal \zeta
\end{equation}
where the sum of the series $\zeta$ is given in~\eqref{eq:zeta}.
Subbing~\eqref{eq:bound_second_half} into~\eqref{eq:bound_first_half} yields the bound~\eqref{eq:state-mal-error-bound-exp}.
    %!TEX root = ../resilient_consensus_trust.tex

\section{Proof of \cref{thm:convergence-rate}}
\label{ass:proof-convergence-rate}

The triangle inequality yields
\begin{equation}\label{eq:convergence-error}
	\begin{aligned}
		\norm[\infty]{\stateleg{t} - \statelegss}	&= \norm[\infty]{\statecontrleg{t} + \statecontrmal{t} - \lr\statecontrlegss + \statecontrmalss\rr}\\
		& \le \norm[\infty]{\statecontrleg{t} - \statecontrlegss} + \norm[\infty]{\statecontrmal{t} - \statecontrmalss}
	\end{aligned}
\end{equation}
where $\statecontrlegss = \one v^\top\statelegtil{}$ and $\statecontrmalss = \one v^\top\statemaltil{}$ represent the final values of legitimate and malicious contributions,
respectively.
We now proceed to bound the two addends in~\eqref{eq:convergence-error},
whereas the uniform bound $2$ in~\eqref{eq:conv-rate-bound} follows from \cref{ass:initial-state-bound}.
    		%!TEX root = ../resilient_consensus_trust.tex

\subsubsection*{Legitimate contribution}
We now prove the bound $\rho_\leg(t)$ in~\eqref{eq:convergence-rate-bound-leg-t>tf}.
For $t\ge\tf$,
it holds
\begin{equation}
	\begin{aligned}
		\statecontrleg{t+1}=\lam{t}\stateleg{0}+(1-\lam{t})\Wlegtrue\statecontrleg{t}.
	\end{aligned}
\end{equation}
Combining this with $v^\top= v^\top \Wlegtrue$ and $a_t\doteq v^\top(\statecontrleg{t}-\stateleg{0})$ yields $a_{t+1}=(1-\lam{t})a_t$ and hence $a_\infty = \prodlamfin{\tf}{\infty}a_{\tf}$,
from which we derive the equality
\begin{equation}\label{eq:decomp-xlegss}
	\statecontrleg{t}-\statecontrlegss = (I-\one v^\top)\statecontrleg{t}+(\prodlamfin{\tf}{t-1}-\prodlamfin{\tf}{\infty})v^\top(\statecontrleg{\tf}-\stateleg{0})\one.
\end{equation}
We bound the norms of the two addends via triangle inequality.
For the first terms, analogously to~\cite[Section~IV]{Blondel05cdc-consensusConvergence},
it holds
\begin{equation}\label{eq:bound-norm-distance-ss}
	\norm[\infty]{(I-\one v^\top)\statecontrleg{t}}
	\le 2\norm[2]{P\statecontrleg{t}}
\end{equation}
with projection matrix $P\in\Real{(L-1)\times L}$ giving $\norm[2]{Px}=\norm[2]{x}$ whenever $x^\top\one=0$.
Apply the triangle inequality to~\eqref{eq:bound-norm-distance-ss}:
\begin{equation}\label{eq:state-mismatch-leg-norm}
	\norm[2]{P\statecontrleg{t}} 
	\le \norm[2]{P\Wautleg{t}\stateleg{0}} 
	+ \norm[2]{P\Winleg{t}\stateleg{0}}.
\end{equation}
We next upper bound these two norms.
For the first,
it holds
\begin{equation}\label{eq:leg-weights-prod-fact-t>t f}
	\begin{aligned}
		\Wautleg{t}
		&= \lr\prodWleg{k}{\tf}{t-1}\rr\prodlamfin{\tf}{t-1} \Wautleg{\tf}\\
		&= \lr\Wlegtrue\rr^{t-\tf} \prodlamfin{\tf}{t-1} \Wautleg{\tf}.
	\end{aligned}
\end{equation}
Let $\one v^\top + VJT$ be a Jordan decomposition of $\Wlegtrue$ where all the eigenvalues in $J\in\Real{(L-1)\times (L-1)}$ are strictly inside the unit circle by \cref{lem:W-primitive}.
Then,
it holds
\begin{align}\label{eq:bound-norm-Waut}
	\begin{split}
		\norm[2]{P\Wautleg{t}\stateleg{0}}
		&= \prodlamfin{\tf}{t-1}\norm[2]{P\lr\Wlegtrue\rr^{t-\tf}\Wautleg{\tf}\stateleg{0}}\\
		&\stackrel{(i)}{=} \prodlamfin{\tf}{t-1}\norm[2]{PVJ^{t-\tf}T\Wautleg{\tf}\stateleg{0}}\\
		&\stackrel{(ii)}{\le}\prodlamfin{\tf}{t-1}\sqrt{L}\norm[\infty]{VJ^{t-\tf}T\Wautleg{\tf}\stateleg{0}}\\
		&\stackrel{(iii)}{\le} \eta \prodlamfin{0}{t-1}\sqrt{L}\norm[\infty]{VJ^{t-\tf}T}\\
		&\stackrel{(iv)}{\le} \eta \frac{b}{2}m\prodlamfin{0}{t-1}\sqrt{L}\binom{t-\tf+m_\specrad}{m_\specrad}\specrad^{t-\tf}
	\end{split}
\end{align}
where $(i)$ uses $P\one = 0$,
$(ii)$ uses $\norm[2]{Px}\le\norm[2]{x}\le \sqrt{L}\norm[\infty]{x}$,
$(iii)$ uses \cref{ass:initial-state-bound} and~\eqref{eq:Waut},
and $(iv)$ follows from powers of Jordan blocks and absorbing $\specrad^{-m_\specrad}$ into $b$\arxiv{, since the largest Jordan block in $J$ has size $m$ and the largest block associated with $\specrad$ has size $m_\specrad$}~\cite{jordan}.
\arxiv{The constant $b$ depends only on $V$ and $T$.}%
We now bound the second addend in~\eqref{eq:state-mismatch-leg-norm}.
We do this analogously to the first addend using the triangle inequality and upper bounding each corresponding summand.
Recalling $\maxof{a}{b}\doteq\max\{a,b\}$,
we rewrite $\Winleg{t}$ as
\begin{equation}\label{eq:B_t-split-tf}
	\begin{aligned}
		\hspace{-2mm}\Winleg{t} &= \sum_{k=0}^{t-1} \prodlamfin{k+1}{t-1}\lr\prodWleg{s}{\maxof{\tf}{(k+1)}}{t-1}\rr\lr\prodWleg{s}{k+1}{\tf-1}\rr\lam{k}\\
		&= \sum_{k=0}^{t-1} \prodlamfin{k+1}{t-1}\lam{k}\lr\Wlegtrue\rr^{t-(\maxof{\tf}{(k+1)})}\lr\prodWleg{s}{k+1}{\tf-1}\rr.
	\end{aligned}
\end{equation}
We use the triangle inequality to bound $\norm[2]{P\Winleg{t}\stateleg{0}}$.
Proceeding analogously to~\eqref{eq:bound-norm-distance-ss}--\eqref{eq:bound-norm-Waut},
we upper bound the $2$-norm of each addend in~\eqref{eq:B_t-split-tf} by the following quantity,
\begin{equation}\label{eq:bound-norm-Win}
	\eta \frac{b}{2}m\sqrt{L}\prodlamfin{k+1}{t-1}\lam{k}\binom{t-(\maxof{\tf}{(k+1)})+m_\specrad}{m_\specrad}\specrad^{t-(\maxof{\tf}{(k+1)})}.
\end{equation}
Combining~\eqref{eq:state-mismatch-leg-norm} with~\eqref{eq:bound-norm-Waut} and~\eqref{eq:bound-norm-Win} yields $\rho_\leg(t)$.
Finally,
for the second addend in~\eqref{eq:decomp-xlegss},
use \cref{ass:initial-state-bound} to get
\begin{equation}\label{eq:bound-xtf}
	\norm[\infty]{v^\top(\statecontrleg{\tf}-\stateleg{0})\one}=|v^\top(\statecontrleg{\tf}-\stateleg{0})|\le 2\eta
\end{equation}
which yields the overall bound in~\eqref{eq:convergence-rate-bound-leg-t>tf}.
    		%!TEX root = ../resilient_consensus_trust.tex

\subsubsection*{Malicious contribution}
We prove the bound $\rho_\mal(t)$ in~\eqref{eq:convergence-rate-bound-mal-t>tf}.
Using~\eqref{eq:Winmal} and $\Wmal{t}\equiv0$ for $t\ge\tfmal$,
the mismatch between state contribution at time $t$ and its final value is
\begin{equation}\label{eq:state-mismatch-mal}
	\begin{aligned}
		\statecontrmal{t} - \statecontrmalss
		&= 
		\sum_{k=0}^{\tfmal-1} \Winmal{k}{t}\statemal{k} - \sum_{k=0}^{\tfmal-1}\Winmal{k}{\infty}\statemal{k}\\
		&=\sum_{k=0}^{\tfmal-1} C_k^{t-1} \prodlamfin{k}{t-1}\Wmal{k}\statemal{k}
	\end{aligned}
\end{equation}
where,
since $k<\tfmal<\tf<\infty$,
\begin{equation}\label{eq:C}
	\begin{aligned}
		C_k^{t-1} &\doteq \prodWleg{s}{k+1}{t-1} - \prodlamfin{t}{\infty}\prodWleg{s}{k+1}{\infty}\\
		&= \lr \lr\Wlegtrue\rr^{t-\tf} - \prodlamfin{t}{\infty}\one v^\top \rr \prodWleg{s}{k+1}{\tf-1}.
	\end{aligned}
\end{equation}
Let us consider the following bound for each summand in~\eqref{eq:state-mismatch-mal},
\begin{equation}\label{eq:contr-mal-prod-bound-t>tf}
	\begin{aligned}
		&\norm[\infty]{C_k^{t-1} \Wmal{k}\statemal{k}} \\
		&=\norm[\infty]{\lr \lr\Wlegtrue\rr^{t-\tfmal} - \prodlamfin{t}{\infty}\one v^\top \rr \prodWleg{s}{k+1}{\tfmal-1}\Wmal{k}\statemal{k}}\\
		&\le\norm[\infty]{\lr\Wlegtrue\rr^{t-\tfmal} - \prodlamfin{t}{\infty}\one v^\top}\norm[\infty]{\prodWleg{s}{k+1}{\tfmal-1}\Wmal{k}\statemal{k}}\\
		&\le\norm[\infty]{\lr\Wlegtrue\rr^{t-\tfmal} - \prodlamfin{t}{\infty}\one v^\top}\norm[\infty]{\Wmal{k}\statemal{k}}.
	\end{aligned}
\end{equation}
Note that applying \cref{lem:matrix-difference} to the matrix difference in~\eqref{eq:contr-mal-prod-bound-t>tf} yields a bound which does not vanish,
hence it is very loose as $t$ gets large.
For any $\tau\ge0$,
it holds by the triangle inequality
\begin{multline}\label{eq:w-true-diff-triangle}
	\norm[1]{\lr\Wlegtrue\rr^{\tau} - \prodlamfin{t}{\infty}\one v^\top}\le 
	\norm[\infty]{\lr\Wlegtrue\rr^{\tau} - \one v^\top} \\
	+ \norm[\infty]{\one v^\top - \prodlamfin{t}{\infty}\one v^\top}.
\end{multline}
The first norm in~\eqref{eq:w-true-diff-triangle} can be bounded in analogy to~\eqref{eq:bound-norm-Waut},
\begin{equation}
	\norm[\infty]{\lr\Wlegtrue\rr^{\tau} - \one v^\top} = \norm[\infty]{VJ^\tau T} \le bm\binom{\tau+m_\specrad}{m_\specrad}\specrad^{\tau}
\end{equation}
while for the second we have
\begin{equation}\label{eq:norm-trivial}
	\norm[\infty]{\one v^\top - \prodlamfin{t}{\infty}\one v^\top} = (1 - \prodlamfin{t}{\infty})\norm[\infty]{\one v^\top} = 1 - \prodlamfin{t}{\infty}.
\end{equation}
\cref{ass:initial-state-bound} and the fact $\max_{i\in\leg}\ls \Wmal{t}\one\rs_{i} \le D_1$ yield
\begin{equation}\label{eq:contr-mal-bound}
	\norm[\infty]{\Wmal{t}\statemal{t}} \le D_1\eta \qquad \forall t\ge0.
\end{equation}
Finally,
applying the triangle inequality to the norm of~\eqref{eq:state-mismatch-mal} with~\eqref{eq:contr-mal-prod-bound-t>tf}--\eqref{eq:norm-trivial} and~\eqref{eq:contr-mal-bound} yields $\rho_\mal(t)$ in~\eqref{eq:convergence-rate-bound-mal-t>tf}.
    \iffalse
    \input{Appendix/proof-convergence-rate-t-less-tf}
    	\input{Appendix/proof-convergence-rate-leg-t-less-tf}	
        \input{Appendix/proof-convergence-rate-mal-t-less-tf}
    \fi		
    %!TEX ROOT = ../resilient_consensus_trust.tex

\section{Ultimate Correct Classification Time}
\label{app:final-time-misclassification-all}

By definition,
time $\tf$ corresponds to ultimate correct classification of malicious and legitimate robots.
Analogously to $\tfmal$ and $\tf$,
there exists finite time $\tfleg$ such that all legitimate robots are detected for $t\ge\tfleg$ almost surely.

%!TEX ROOT = ../resilient_consensus_trust.tex

\subsubsection*{Ultimate classification of malicious robots}

We are concerned with the following joint probability:
\begin{equation}\label{eq:final-time-misclassification-mal-robots-def}
	\pr{\tfmal=k} = \pr{\eventcorrmal{t} \,\forall t\ge k \wedge \eventmisclmal{k-1}}
\end{equation}
where the events $\eventcorrmal{t}$ and $\eventmisclmal{t}$,
respectively corresponding to correct classification of all malicious robots at time $t$ and misclassification of (at least) one malicious robot at time $t$,
are defined as
\begin{align}
	\eventcorrmal{t} &\doteq \lb\trusthist{i}{j}{t}<0 \, \forall i\in\leg,j\in\mal\rb \label{eq:event-correct-classification-mal}\\
	\eventmisclmal{t} &\doteq \lb\exists i\in\leg,j\in\mal : \trusthist{i}{j}{t}\ge0\rb. \label{eq:event-misclassification-mal}
\end{align}
From marginalization,
it follows
\begin{equation}\label{eq:final-time-misclassification-mal-robots-1}
	\pr{\eventcorrmal{t} \,\forall t\ge k \wedge \eventmisclmal{k-1}} \le \pr{\eventmisclmal{k-1}}.
\end{equation}
Applying the union bound yields
\begin{equation}\label{eq:final-time-misclassification-mal-robots-2}
	\begin{aligned}
		\pr{\eventmisclmal{k-1}}	&\le \sum_{i\in\leg}\sum_{j\in\mal\cap\neigh{i}} \pr{\trusthist{i}{j}{k-1}\ge0}\\
		&\stackrel{(i)}{\le} \sum_{i\in\leg}\sum_{j\in\mal\cap\neigh{i}} \e^{-2\meanmal^2k}\\
		&\stackrel{(ii)}{=}D_\mal\e^{-2\meanmal^2k}
	\end{aligned}
\end{equation}
where~\eqref{eq:prob-misclassification} is used in $(i)$
and~\eqref{eq:D} is used in $(ii)$.
Combining~\eqref{eq:final-time-misclassification-mal-robots-def},~\eqref{eq:final-time-misclassification-mal-robots-1} and~\eqref{eq:final-time-misclassification-mal-robots-2} yields
\begin{equation}
	\pr{\tfmal=k} \le D_\mal\e^{-2\meanmal^2 k}, \qquad k\ge 0.
\end{equation}

\arxiv{Moreover,
a tighter bound,
which is however more difficult to use in analysis,
can be derived as follows:
\begin{equation}\label{eq:final-time-misclassification-mal-robots-tighter-bound}
	\begin{aligned}
		\pr{\eventmisclmal{k-1}} &=1 - \pr{\eventcorrmal{k-1}}\\
		&=1 - \pr{\trusthist{i}{j}{k-1} < 0 \, \forall i\in\leg,\forall j\in\mal}\\
		&=1 - \prod_{i\in\leg} \prod_{j\in\mal} \pr{\trusthist{i}{j}{k-1} < 0}\\
		&=1 - \prod_{i\in\leg} \prod_{j\in\mal} \lr 1 - \pr{\trusthist{i}{j}{k-1} \ge0} \rr\\
		&\le 1 - \prod_{i\in\leg} \prod_{j\in\mal} \lr 1 - \e^{-2k\meanmal^2}\rr\\
		&=1 - \prod_{i\in\leg} \lr 1 - \e^{-2k\meanmal^2}\rr^{|\mal\cap\neigh{i}|}\\
		&=1 - \lr 1 - \e^{-2k\meanmal^2}\rr^D.
	\end{aligned}
\end{equation}}
%!TEX ROOT = ../resilient_consensus_trust.tex

\subsubsection*{Ultimate classification of legitimate robots}

We now address the probability
\begin{equation}\label{eq:final-time-misclassification-leg-robots-def}
	\pr{\tfleg=k} = \pr{\eventcorrleg{t} \,\forall t\ge k \wedge \eventmisclleg{k-1}}
\end{equation}
where the events $\eventcorrleg{t}$ and $\eventmisclleg{t}$,
respectively corresponding to correct classification of all legitimate robots at time $t$ and misclassification of one legitimate robot at time $t$,
are
\begin{align}
	\eventcorrleg{t} &\doteq \lb\trusthist{i}{j}{t}\ge0 \, \forall i\in\leg,j\in\leg\rb \label{eq:event-correct-classification-leg}\\
	\eventmisclleg{t} &\doteq \lb\exists i\in\leg,j\in\leg : \trusthist{i}{j}{t}<0\rb. \label{eq:event-misclassification-leg}
\end{align}
Analogously to classification of malicious robots,
applying marginalization and the union bound to~\eqref{eq:final-time-misclassification-leg-robots-def} yields
\begin{equation}\label{eq:final-time-misclassification-leg-robots-1}
	\pr{\tfleg=k} \le D_\leg\e^{-2\meanleg^2k}, \qquad k\ge 0.
\end{equation}
\arxiv{Moreover,
a tighter bound can be derived akin~\eqref{eq:final-time-misclassification-mal-robots-tighter-bound}:
\begin{equation}\label{eq:final-time-misclassification-leg-robots-tighter-bound}
	\begin{aligned}
		\pr{\eventmisclleg{k-1}} \le 1 - \lr 1 - \e^{-2k\meanleg^2}\rr^{D_2}.
	\end{aligned}
\end{equation}}

\subsubsection*{Ultimate classification time}
Applying the union bound to all events considered in the previous two cases readily yields
\begin{equation}
	\pr{\tf=k} \le D_\leg\e^{-2\meanleg^2k} + D_\mal\e^{-2\meanmal^2k}, \quad k\ge 0\journalVersion{}{.}
\end{equation}
\arxiv{and
\begin{equation}
	\pr{\tf=k} \le 2 - \lr 1 - \e^{-2k\meanmal^2}\rr^D_\mal - \lr 1 - \e^{-2k\meanleg^2}\rr^{D_\leg}.
\end{equation}}
    \arxiv{%!TEX ROOT = ../resilient_consensus_trust.tex

\section{Tighter Bound for Deviation due to Malicious Agents}
\label{app:tighter-bound-mal-agents}

Let us first note that
\begin{equation}
	D_1 = \dfrac{\max_{i\in\leg}|\mal\cap\neigh{i}|}{\max_{i\in\leg}|\mal\cap\neigh{i}| + 1}
\end{equation}
where the equality follows because $\frac{n}{n+1}$ is increasing with $n$.
The weights given to malicious agents are upper bounded as
\begin{equation}
	\begin{aligned}
		\ls \Wmal{t}\one\rs_{i} = \sum_{j=1}^M \ls \Wmal{t}\rs_{ij}	&=\dfrac{\sum_{j\in\mal}\one[\beta_{ij}(t)\ge0]}{|\neigh[t]{i}|+1}\\
		&\le \frac{\sum_{j\in\mal}\one[\beta_{ij}(t)\ge0]}{\sum_{j\in\mal}\one[\beta_{ij}(t)\ge0]+1}\\
		&\stackrel{(i)}{\le}  \dfrac{|\mal\cap\neigh{i}|}{|\mal\cap\neigh{i}| + 1}
	\end{aligned}
\end{equation}
where $(i)$ follows because $\frac{n}{n+1}$ is increasing with $n$ and it holds for every $i\in\leg$
\begin{equation}
	\sum_{j\in\mal}\one[\beta_{ij}(t)\ge0] \le \sum_{j\in\mal} 1 = |\mal\cap\neigh{i}| = D_\mal.
\end{equation}
Using~\eqref{eq:D1},
we can tighten bound~\eqref{eq:weight-mal-upper-bound-max} as follows:
\begin{equation}\label{eq:weights-mal-mean-upper-bound-min}
	\mean{\max_{i\in\mathcal{L}}\ls \Wmal{t}\one\rs_{i}} \le \min\lb D_1, \dfrac{D_\mal}{2}\e^{-2(t+1)\meanmal^2}\rb.
\end{equation}
By defining the threshold time instant $\bar{k}_1$ as
\begin{equation}
	\bar{k}_1 \doteq \floor*{\dfrac{1}{2\meanmal^2}\log\frac{D_\mal}{2D_1}},
\end{equation}
the bound~\eqref{eq:weights-mal-mean-upper-bound-min} can be equivalently expressed as
\begin{equation}\label{eq:weights-mal-mean-upper-bound-cases}
	\mean{\max_{i\in\mathcal{L}}\ls \Wmal{t}\one\rs_{i}} \le 
	\begin{cases}
		D_1, & t \le \bar{k}_1\\
		\dfrac{D_\mal}{2}\e^{-2(t+1)\meanmal^2}, & t > \bar{k}_1.
	\end{cases}
\end{equation}
Then,
the upper bound~\eqref{eq:state-mal-error-mean-bound} can be refined as
\begin{equation}\label{eq:state-mal-error-mean-bound-tighter}
	\begin{aligned}
		\mean{\statemalerr{i}{t}}	&\le %\mean{\dfrac{\eta}{2} \sum_{k=0}^{\tfmal-1} (1-\lam{k+1}) (1-\lam{k}) \max_{i\in\mathcal{L}}\ls \Wmal{t}\one\rs_{i}}\\
																\dfrac{\eta}{2} \sum_{k=0}^{\tfmal[t]-1} (1-\lam{k+1}) (1-\lam{k}) \mean{\max_{i\in\mathcal{L}}\ls \Wmal{t}\one\rs_{i}}\\
		&\stackrel{(i)}{\le} \dfrac{\eta}{2} \lr S_1(\minof{(\tfmal[t]-1)}{\bar{k}_1}) + S_2(\tfmal[t])\rr
	\end{aligned}
\end{equation}
where~\eqref{eq:weights-mal-mean-upper-bound-cases} is used in $(i)$ and
\begin{align}
	S_1(t) &= D_1\sum_{k=0}^{t} (1-\lam{k+1}) (1-\lam{k})\\
	S_2(t) &= \dfrac{D_\mal}{2}\sum_{k=\bar{k}_1+1}^{t-1} (1-\lam{k+1}) (1-\lam{k})\e^{-2(k+1)\meanmal^2}.
\end{align}
The probability of final correct classification time of malicious agents is upper bounded in~\cref{app:final-time-misclassification-all} as
\begin{equation}\label{eq:prob-misclassification-mal-agents-min}
	\pr{\tfmal = k} \le D_\mal\e^{-2k\meanmal^2}.
\end{equation}
By defining the threshold time instant $\bar{k}_2$ as
\begin{equation}
	\bar{k}_2 \doteq \floor*{\dfrac{\log D_\mal}{2\meanmal^2}},
\end{equation}
the bound~\eqref{eq:prob-misclassification-mal-agents-min} can be equivalently expressed as
\begin{equation}\label{eq:prob}
	\pr{\tfmal = k} \le 
	\begin{cases}
		1, & k \le \bar{k}_2\\
		D_\mal\e^{-2k\meanmal^2} & k > \bar{k}_2.
	\end{cases}
\end{equation}
Clearly $\bar{k}_2<\bar{k}_1$.
Putting everything together,
the total bound on deviation due to malicious agents becomes
\begin{equation}\label{eq:bound-mal-deviation-tighter}
	\begin{aligned}
		\mean{\lim_{t\rightarrow\infty}\statemalerr{i}{t}}	&\le\begin{multlined}[t]
			\sum_{k=0}^{\infty}\dfrac{\eta}{2} S_1(\minof{(k-1)}{\bar{k}_1})\pr{\tfmal=k} \\
			+\sum_{k=0}^{\infty}\dfrac{\eta}{2}  S_2(k)\one[k>\bar{k}_1+1]\pr{\tfmal=k}
		\end{multlined}\\
															&= \begin{multlined}[t]
																\dfrac{\eta}{2}\sum_{k=0}^{\bar{k}_1} S_1(k)\pr{\tfmal=k} \\
																+\dfrac{\eta}{2}\sum_{k=\bar{k}_1+1}^{\infty} \lr S_1(\bar{k}_1) + S_2(k)\rr\pr{\tfmal=k}
															\end{multlined}\\
															&\le\begin{multlined}[t]
																\dfrac{\eta}{2}\sum_{k=0}^{\bar{k}_2} S_1(k) +\dfrac{D_\mal\eta}{2}\sum_{k=\bar{k}_2+1}^{\bar{k}_1} S_1(k) \e^{-2k\meanmal^2}\\
																+\dfrac{D_\mal\eta}{2}\sum_{k=\bar{k}_2+1}^{\infty} \lr S_1(\bar{k}_1) + S_2(k)\rr\e^{-2k\meanmal^2}.
															\end{multlined}
	\end{aligned}
\end{equation}
Both the summations and the sum of the series in~\eqref{eq:bound-mal-deviation-tighter} can be computed exactly through formulas for geometric sequences.

\begin{comment}
	\begin{figure}
		\centering
		\includegraphics[width=\linewidth]{Images/polynomial_decay.png}
		\caption{Polynomially decaying lambdas
		}
		\label{fig:deviation-poly}
	\end{figure}
	\begin{figure}
		\centering
		\includegraphics[width=\linewidth]{Images/exponential_decay.png}
		\caption{Exponentially decaying lambdas
		}
		\label{fig:deviation-exp}
	\end{figure}
\end{comment}}
 

%    \bibliographystyle{IEEEtran}
%	\bibliography{Bibfiles/bibfile, 
%    	Bibfiles/biboptions, 
%    	IEEEabrv
%    }
	
\end{document}